\documentclass[twocolumn]{aastex7}

\newcommand{\mdiff}{\mathrm{d}}
\newcommand{\amin}{$^{\prime}$}
\newcommand{\asec}{$^{\prime\prime}$}
\newcommand{\kms}{\,km\,s$^{-1}$}
\newcommand{\um}{\,$\mu$m}
\newcommand{\mdeg} {^\circ}

\newcommand{\cpl} {C$^{+}$}

\newcommand{\oi} {{\sc [O\,i]}} 
 
\newcommand{\oiii}{{\sc [O\,iii]}} 
\newcommand{\oiv}{{\sc [O\,iv]}} 
 
\newcommand{\cii}{{\sc [C\,ii]}} 
\newcommand{\nii}{{\sc [N\,ii]}} 
\newcommand{\niii}{{\sc [N\,iii]}} 
 
\newcommand{\hii}{{\sc H\,ii}} 
\newcommand{\htwo}{H$_2$} 
\newcommand{\oisw}{\mbox{{\sc [O\,i]}\,63\,$\mu$m}} 
\newcommand{\oilw}{\mbox{{\sc [O\,i]}\,145\,$\mu$m}}

\newcommand{\ciiw}{\mbox{{\sc [C\,ii]}\,158\,$\mu$m}}

\newcommand{\thco}{$^{13}$C$^{}$O}
\newcommand{\ceio}{$^{}$C$^{18}$O}

\newcommand {\ra}[3]{$\alpha_{J2000} = #1^{\rm h} #2^{\rm m} #3^{\rm s}$} 
\newcommand {\dec}[3]{$\delta_{J2000} = #1^\circ #2^\prime #3^{\prime\prime}$} 

\newcommand \persqcm {\,cm$^{-2}$}
\newcommand \percucm {\,cm$^{-3}$}

\newcommand \sgra {Sgr\,A}
\newcommand \sgrae {Sgr\,A East}
\newcommand \sgraw {Sgr\,A West}
\newcommand \sgras {Sgr\,A*}
\newcommand \sgrb {Sgr\,B}
\newcommand \sgrc {Sgr\,C}

\newcommand \halo {Radio Arc Halo}

\newcommand \grt {upGREAT}

\newcommand \pell {$+\ell$}
\newcommand \mell {$-\ell$}

\usepackage{enumitem}
\begin{document}

\title{SOFIA/\grt\ imaging spectroscopy of the \ciiw\ fine structure
  line \\toward the Sgr\,A region in the Galactic center}

\shorttitle{\cii\ toward \sgra}
\shortauthors{Harris et al.}
\correspondingauthor{Andrew Harris}
\email{harris@astro.umd.edu}

\author[orcid=0000-0001-6159-9174]{A.I.~Harris} 
\affil{Department of Astronomy, University of Maryland, College Park,
  MD 20742, USA}
\email{harris@astro.umd.edu}

\author[orcid=0000-0002-1708-9289]{R.~G\"usten} 
\affiliation{Max-Planck-Institut f\"ur Radioastronomie, Auf dem
  H\"ugel 69, 53121 Bonn, Germany}
\email{guesten@mpifr-bonn.mpg.de}

\author{M.A.~Requena-Torres}
\affiliation{Department of Physics, Astronomy,
  and Geosciences, Towson University, Towson, MD 21252, USA}
\email{mrequenatorres@towson.edu}

\author[orcid=0000-0001-5389-0535]{D.~Riquelme}
\affiliation{Max-Planck-Institut f\"ur
  Radioastronomie, Auf dem H\"ugel 69, 53121 Bonn, Germany}
\affiliation{Departamento de Astronom\'ia, Universidad de
  La Serena, Ra\'ul Bitr\'an 1305, La Serena, Chile}
\email{denise.riquelme@userena.cl}

\author[orcid=0000-0002-6753-2066]{M.R.~Morris}
\affiliation{Department of Physics and Astronomy, University of
  California, Los Angeles, CA 90095, USA}
\email{morris@astro.ucla.edu}

\author{G.J.~Stacey} 
\affiliation{Department of Astronomy, Cornell University, Ithaca, NY
  14853, USA}
\email{stacey@astro.cornell.edu}

\author[orcid=0000-0001-7658-4397]{J.~Stutzki} 
\affiliation{I. Physikalisches Institut der Universit\"at zu K\"oln,
  Z\"ulpicher Stra{\ss}e 77, 50937 K\"oln, Germany}
\email{stutzki@ph1.uni-koeln.de}

\author[orcid=0000-0002-6838-6435]{Y.~Okada} 
\affiliation{I. Physikalisches Institut der Universit\"at zu K\"oln,
  Z\"ulpicher Stra{\ss}e 77, 50937 K\"oln, Germany}
\email{okada@ph1.uni-koeln.de}

\author[orcid=0000-0003-4195-1032]{E.~Chambers} 
\affiliation{SOFIA Science Center, Universities Space Research
  Association, NASA Ames Research Center, Moffett Field, CA 94035,
  USA}
\affiliation{Space Science Institute, 4765 Walnut St, Suite B,
  Boulder, CO 80301, USA}
\email{echambers@spacescience.org}

\author{M.~Mertens}
\affiliation{I. Physikalisches Institut der Universit\"at zu K\"oln,
  Z\"ulpicher Stra{\ss}e 77, 50937 K\"oln, Germany}
\affiliation{Max-Planck-Institut f\"ur Radioastronomie, Auf dem
  H\"ugel 69, 53121 Bonn, Germany}
\email{mmertens@mpifr-bonn.mpg.de}

\author[orcid=0000-0002-7299-8661]{C.~Fischer} 
\affiliation{Deutsches SOFIA Institut, University of Stuttgart,
  70569 Stuttgart, Germany}
\email{fischer@dsi.uni-stuttgart.de}

\begin{abstract}
  We present SOFIA/\grt\ velocity-resolved spectral imaging and
  analysis of the $\lambda 158$\um\ \cii\ spectral line toward the
  central 80 by 43\,pc region of the Central Molecular Zone of the
  Galaxy.  The field we imaged with 14\arcsec\ (0.6\,pc) spatial and
  1\kms\ spectral resolution contains the Circum-Nuclear Disk (CND)
  around the central black hole \sgras, the neighboring thermal Arched
  Filaments, the nonthermal filaments of the Radio Arc, and the three
  luminous central star clusters.  \cii\ traces emission from the
  CND's inner edge to material orbiting at a distance of approximately
  6\,pc.  Its velocity field reveals no sign of inflowing material nor
  interaction with winds from the \sgrae\ supernova
  remnant. Wide-field imaging of the \sgra\ region shows multiple
  circular segments, including the thermal Arched Filaments, that are
  centered on a region that includes the Quintuplet cluster.  We
  examine the possibility that the Arched Filaments and other
  large-scale arcs trace transient excitation events from supernova
  blast waves.  Along the Arched Filaments, comparisons among far-IR
  fine structure lines show changes in ionization state over small
  scales and that high-excitation lines are systematically shifted in
  position from the other lines.  These also point to transient fast
  winds that shocked on the surface of the Arches cloud to produce
  additional local UV radiation to excite the Arched Filaments on a
  cloud surface illuminated by UV from hot stars.
\end{abstract}



\section{Introduction} \label{sec:intro}

Our Galactic center is both a unique part of the Galaxy and an analog
of other ``normal'' spiral galactic nuclei.  Understanding its
structure, physical conditions, kinematics, dynamics, and other
properties gives us a comprehensive picture of a spiral galactic
nucleus with moderate activity.  Some of the many reviews of the
region are \citet{brown84}, \citet{morris96}, \citet{genzel10}, and
\citet{henshaw23}.  Detailed studies allow us to identify sources of
luminosity and separate which structures are large-scale and could be
common to many nuclei (e.g., flows along bars, orbit crowding,
interactions between molecular clouds and magnetic fields) and which
are transient (e.g., starburst clusters, cloud-cloud collisions, or
supernovae shells).

Unraveling the structure and kinematics of this complex region
requires continuum and spectral line observations at many wavelengths.
Observations at infrared and longer wavelengths are necessary because
material in the intervening Galactic plane blocks our view in the
visible and ultraviolet.  The center's proximity enables detailed
studies, even from telescopes with moderate aperture, since 1\arcsec\
corresponds to about 0.04\,pc at the center.

\cii\ emission provides one of the clearest views of the Galactic
center.  With an 11.3\,eV excitation potential, \cii\ traces both
ionized material in \hii\ regions and atomic gas in photodissociation
regions (PDRs), where it is a major coolant (e.g., \citealt{tielens85,
  crawford85, rubin85, vandishoeck88, wolfire90, stacey91,
  sternberg95, kaufman99, goldsmith12, langer14, pineda14,
  herreracamus15}).  Its 158\um\ wavelength penetrates much of the
dust obscuration from line-of-sight molecular clouds.  Typical
molecular clouds in the Galactic disk cause deep \cii\ absorption
features along the line of sight to the center, but emit only weakly,
since few clouds in the plane have the strong UV from young stars
necessary for intense \cii\ emission.

\cii\ traces material excited by UV from stars and shocks; both are
prevalent in the Galactic center. Velocity information from
high-resolution spectroscopy separates absorption and emission
features to pick apart structures along the line of
sight. Velocity-resolved spectroscopy together with a kinematic model
adds the third dimension for understanding the structure of the
Galactic nuclear region.

Here we present spectral imaging and analysis of \cii\ from the
central field of our large SOFIA/\grt\ program to image the Central
Molecular Zone of the Galaxy in \cii\ \citep{wholecmz}.  Among other
features, this field contains the Circum-Nuclear Disk (CND) around the
central black hole \sgras, the neighboring thermal Arched Filaments
that stand out in radio continuum and warm dust emission, the Sickle
and Pistol regions, and the Quintuplet, Arches, and Central Nuclear
star clusters.

Naming conventions in this paper closely follow the names derived from
radio continuum measurements at a range of angular resolutions.  At
modest resolution, the Galactic center's main luminosity is associated
with bright radio continuum from the \sgra, \sgrb, \sgrc\ regions.

Data in this paper cover the ``\sgra\ region'' that encompasses the
most luminous parts of \sgra\ and surroundings.  The brightest radio
continuum is associated with the \sgra\ source at the center of the
Galaxy.  This includes the \sgraw\ and \sgrae\ sources: the former
consists of the \sgras\ point source and its surrounding molecular
Circum-Nuclear Disk (CND) and radio continuum ``mini-spiral.''
\sgraw\ lies toward the western edge of the $2^\prime \times 3^\prime$
\sgrae\ supernova remnant shell.  Both are surrounded by a \sgra\
radio continuum halo that is about 7\amin\ in diameter. (We use the
descriptive term halo to describe approximately circular radio
continuum emission regions that do not show the limb-brightening
characteristic of a hollow shell or bubble.)
  
A 10\arcmin\ diameter radio continuum halo lies some 14\amin\ to
positive Galactic longitude ($+\ell$, northeast) from \sgras.  It is
bisected by the nonthermal filaments that constitute the Radio Arc and
span its diameter.  Due to its size, interferometers resolve out much
of the halo's flux, so it is more prominent in single-dish images.  We
use the name \halo\ because it is descriptive and avoids the ambiguity
of the name ``Radio Arc Bubble,'' which has been used both for this
halo (e.g., \citealt{heywood22}) and another, smaller, region also
known as the MSX Bubble or the Arc Bubble (e.g., \citealt{rodriguez01,
  simpson07}).  We use the term MSX bubble for that approximately
circular bright rim with radius 190\arcsec\ centered at Galactic
coordinates $(\ell, b) \sim (0.13\mdeg, -0.1\mdeg)$. The rim surrounds
an infrared-dark region that is filled with X-ray and high-excitation
mid-IR line flux \citealt{egan98, price01, rodriguez01, simpson07,
  ponti15, molinari16}).

The Arches region lies about halfway between \sgras\ and the center of
the \halo, and is above the Galactic plane.  We use the term ``Arches
region'' to include the thermal Arched Filaments and the background
molecular cloud at the same velocity.

After summarizing the observations and data reduction in
Section~\ref{sec:obs}, we describe the \cii\ distribution and velocity
field in Section~\ref{sec:results}, putting them in context with other
probes of the material in the center.  Our discussion in section
\ref{sec:discuss} contains a general overview of what the \cii\
observations reveal or suggest, then focuses on the analysis of the
area around the CND (Sec.~\ref{sec:cnd}) and the Arched Filaments
(Sec.~\ref{sec:circles}).  Section~\ref{summary} is a brief summary of
our main findings and speculations.

In calculations, we use a single representative distance for all
objects in the \sgra\ region equal to the \sgras\ distance of 8.2\,kpc
\citep{gravity19}.
 
\section{Observations} \label{sec:obs}

\subsection{SOFIA/\grt}
Our \grt\footnote{\grt\ is a development by the Max-Planck-Institut
  f\"ur Radioastronomie and the I.~Physikalisches Institut of the
  Universit\"at zu K\"oln, in cooperation with the DLR Institut f\"ur
  Optische Sensorsysteme.} \citep{risacher18} observations on the
Stratospheric Observatory For Infrared Astronomy (SOFIA,
\citealt{young12}) have been described in \citet{harris21}, so we
provide only a brief summary here.  Additional technical details are
available in \citet{sgrc} and \citet{wholecmz}.  

As before, the basic sampling strategy for the entire project
\citep{wholecmz} was to observe individual spectral cubes, each of
which covered $560 \times 560$ arcseconds in Galactic longitude $\ell$
and latitude $b$ over LSR velocities $-190$ to $+220$\kms\ in the
1.901\,THz ($\lambda$157.74\um) \cii\
$\mathrm{^2P_{3/2}} - \mathrm{^2P_{1/2}}$ fine structure spectral
line.  In this paper we show and discuss the $4 \times 2$ spectral
cubes that covers a the \sgra\ region with area
$\Delta\ell = 0.557\degr$ by $\Delta b = 0.303\degr$ centered at
($\ell, b$) = ($0.133\degr, -0.052\degr$) (\ra{17}{46}{08},
\dec{-28}{58}{07}).  This region abuts the \sgrb\ region of
\citet{harris21}.  All data were obtained in observing campaigns
flying from New Zealand in 2017 June and July and 2018 June in
programs 05\_0022, 06\_0173, and 83\_0609.

We used \grt's dual frequency on-the-fly mapping mode with 7-pixel
arrays of hot electron bolometer mixers and Fast Fourier Transform
Spectrometers (FFTS4G, updated from \citealt{klein12}).  The Low
Frequency Array was tuned for the \cii\ line, providing a main beam
FWHM of 14.1\asec.  The rms pointing accuracy was 2\asec.  Amplitude
calibration is natively on a Rayleigh-Jeans brightness temperature
scale as a function of velocity $T_B(\mathrm{v})$, corresponding to
intensities $I$ through
\begin{equation}
I = \int I_\nu (\nu) \, \mdiff\nu = \frac{2k}{\lambda^3} 
\int T_B(\mathrm{v}) \, \mdiff{\rm v} \;.
\label{eq:intens}
\end{equation}
Estimated absolute intensity uncertainties are 20\%.  Off-line
baseline structure was removed by fitting with baseline structures
derived from differences between nearby ``off'' spectra
\citep{higgins11, kester14, higgins21}.  Efficient on-the-fly imaging
required a ``near'' spectral reference position (\ra{17}{47}{41.3},
\dec{-28}{35}{00}) and a ``far'' reference position
(\ra{17}{55}{03.9}, \dec{-29}{23}{02}) to measure and remove residual
\cii\ emission in the near position.

\subsection{{\it Herschel}/PACS}
In our discussion section~\ref{sec:circles} we use some complementary
results from observations in our {\em Herschel} EXtraGalactic (HEXGAL)
guaranteed time key project (P.I.: R.~G\"usten, proposal name {\em
  KPGT\_rguesten\_1}). {\em Herschel}-PACS \citep{pilbratt10,
  poglitsch10} mapped the $\lambda 122$\,\um\ \nii\ and
$\lambda 145$\,\um\ \oi\ lines with its Red channel on 2010 October
18, and $\lambda 63$\,\um\ \oi\ and $\lambda 88$\,\um\ \oiii\ lines
with its Blue channel on 2011 April 05.  We took instrumental
characteristics from the \citet{pacsman} and \citet{poglitsch10}.  The
spatial resolution was approximately 9.4\asec\ at 63\um\ and 88\um,
10\asec\ at 121\um, and 11\asec\ at 145\um.  The spectral resolution
was 100--120\kms\ FWHM at the shorter wavelengths and 250--300\kms\ at
the longer.  We used the PACS unchopped spectroscopy mode with a
distant ``off'' position of \ra{17}{44}{33.5}, \dec{-28}{52}{08.3},
approximately 16\arcmin\ from the Arches region and 14\arcmin\ off the
Galactic plane, to measure the telescope background spectrum for the
observations.  This mode was essential for observations within the
center's bright extended emission.

We used HIPE\,11 for initial reductions, then wrote data to FITS files
for exploratory imaging and final processing with other software
tools.  Inspection of data from all spaxels showed that the lines were
unresolved within the spectrometer's resolution.  We measured the
integrated intensity of the unresolved lines by summing over spectral
bins containing intensity in the filter core and wings after
subtracting a linear baseline determined by off-line spectral bins to
either side of the line bins.  In spot comparisons, the PACS
intensities agree within a factor of two with values reported along E2
and toward the G\,0.095+0.012 \hii\ region by \citet{erickson91}.  We
use the HIPE\,11 values in this paper.

\subsection{SOFIA/FIFI-LS}
We show complementary \niii\ ($\lambda 57$\um) data in
Section~\ref{sec:circles}.  These data were taken in SOFIA program
04\_0032 on 2016 June 30 and July 6 during flights from Christchurch,
New Zealand in the blue channel of FIFI-LS \citep{fischer18}.
Parallel \cii\ observations in the red channel will be published
elsewhere in reports of the instrument team's guaranteed time
observations. The observing mode was chop-nod pairs with an
asymmetrical 590\asec\ total chop throw at position angle $123\mdeg$
east of north (June 30) and $0\mdeg$ (July 6).  Residual telescope
emission was removed with data at a reference position 21\arcmin\ to
the northwest relative to each field. FIFI-LS data were reduced with
the instrument's standard pipeline \citep{fadda23,
  fifical20}. Telluric atmosphere corrections are from
satellite-derived water vapor values calibrated to direct measurements
of the water vapor with FIFI-LS \citep{fischer21, iserlohe21}. Spectra
had a zeroth-order polynomial baseline removed, with intensities
obtained from Gaussian profile fits to the line emission in 6\asec\
diameter apertures around each spatial pixel to match the angular
resolution of the SOFIA telescope.  The FWHM velocity resolution was
290\kms.

\section{Results}\label{sec:results}
\subsection{Spatial distributions}\label{ssec:large}
\begin{figure*}[p]
\centering
\includegraphics[width=0.7\textwidth, trim=130 120 203 160, clip]{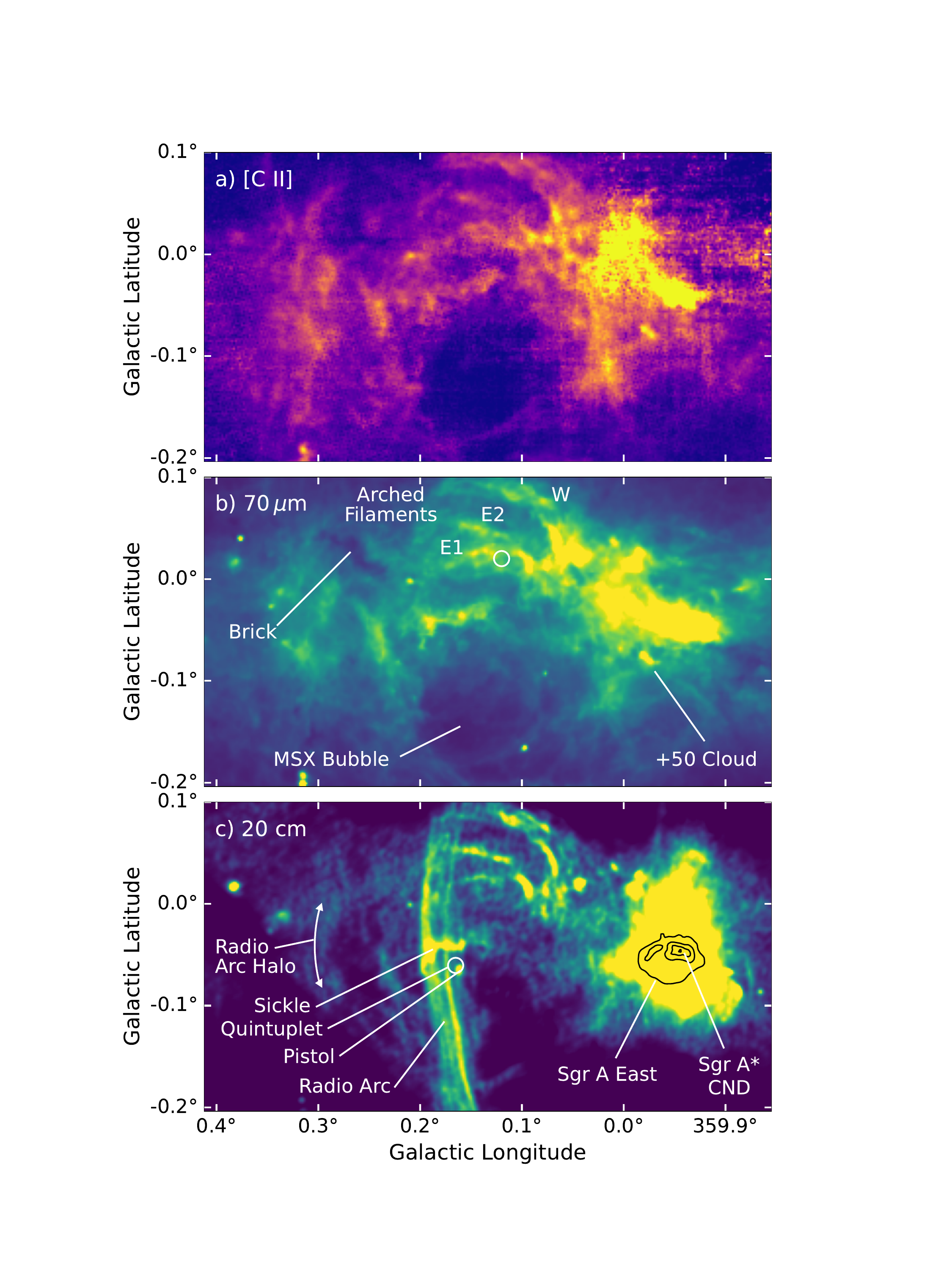}
\caption{Overviews for comparisons.  Top: \cii\ integrated intensity
  over $\pm 120$\kms\ (intensity scale in Fig.~\ref{fig:moments}).
  Middle: 70\um\ continuum emission \citep{molinari16}.  Bottom:
  20\,cm radio continuum \citep{lang10}.  Images are on linear scales
  with the brightest regions saturated to show extended
  emission. Labels in the 70\um\ and 20\,cm images highlight notable
  regions including dust continuum from the W, E2, and E1 Arched
  Filaments; the MSX Bubble; the Brick; the nonthermal filaments that
  form the Radio Arc at $\ell \approx 0.18\mdeg$; an arc to indicate a
  section of the outer edge of the \halo; and contours tracing
  emission from \sgras\ and the Circum-Nuclear Disk at the western
  edge of the \sgrae\ nonthermal radio bubble. Open circles indicate
  the positions of the Arches cluster near the E1 filament in panel b)
  and Quintuplet cluster, near the Sickle in panel
  c). \label{fig:overview3labels}}
\end{figure*}

Our spectroscopic imaging data provide a much more complete view of
the central region in \cii\ than previous maps, cuts, or observations
of individual regions within the field provided (e.g.,\ \citealt{genzel90,
  poglitsch91, mizutani94, garcia16, langer17}).
Figure~\ref{fig:overview3labels} provides orientation for the \sgra\
region.  The top panel is our \cii\ integrated intensity image
covering ${\rm v}_{\rm LSR} = \pm120$\kms\ and all spectral channels
with signal $\geqslant 3\sigma$.  The middle and bottom panels of the
figure show the 70\um\ {\it Herschel}/PACS and 20\,cm radio continuum
images \citep{molinari16, lang10} for comparison.  The images are on
linear scales, truncated at the highest intensities to more clearly
show the extended emission.

For orientation, the 20\,cm\ and 70\um\ panels have labels identifying
some of the Galactic center's prominent features.  Contours on the
20\,cm image show the \sgra\ complex, which is composed of \sgrae\ and
\sgraw\ (e.g., \citealt{ekers83, yusefzadeh87, pedlar89}). \sgrae\ is
a roughly elliptical region filled with nonthermal radio emission,
with \sgraw\ at its western edge.  \sgraw\ contains the \sgras\ black
hole and its surrounding ``mini-spiral'' delineating the inner edge of
the Circum-Nuclear Disk (CND).  The W, E2, and E1 Arched Filaments
\citep{morris89} near $\ell \approx 0.1\mdeg$ are visible in all
panels.  The Sickle and Pistol regions lie near the nonthermal
filaments of the Radio Arc detected in radio continuum synchrotron
emission.  Two open circles mark the positions of the Quintuplet and
Arches clusters of hot young stars \citep{glass90, cotera96, figer99a,
  figer99b}; the third massive star cluster is the Milky Way Central
Nuclear Cluster centered on \sgras\ (e.g., \citealt{becklin68,
  schoedel14}).

The distributions of extended emission traced in \cii\ and 70\um\ are
very similar, notably toward the Arched Filaments and in the region of
extended bright emission extending from $\ell \approx 0.2\mdeg$ to
0.4\degr.  Whether originating in PDRs or \hii\ regions, \cii\
requires UV for excitation, with the photons also heating dust that
emits at 70\um.  The good spatial correspondence between \cii\ and
70\um\ in this region and in the Arched Filaments point to a common
origin related to UV illumination across the entire region.  With the
exception of the nonthermal filaments of the Radio Arc, there is
general commonality between \cii\ and the extended 20\,cm thermal
continuum across the region as well. There is little correspondence
between the more uniform 160\um\ continuum \citep{molinari16} or CO,
\thco, and \ceio\ integrated intensity distributions \citep{apexco},
however.  Good spatial correspondence at 70\um\ and 20\,cm continua,
but poor correspondence in tracers of molecular cloud column
densities, strongly indicates that \cii\ traces UV at cloud surfaces
rather than the cloud bodies.

The close spatial agreement between the \cii\ integrated intensity and
70\um\ spatial distributions in Fig.~\ref {fig:overview3labels} is
evidence that Galactic absorption masks only a small amount of their
emission from the Galactic center.  Absorption from the ``Brick,'' a
very dense cloud with little active star formation (M\,0.25+0.011,
G\,0.253+0.016, \citealt{guesten81, lis94, longmore12, immer12}), is
visible as a dark patch in the 70\um\ and \cii\ images.  The Brick has
no evident 20\,cm emission, so none of its surfaces are strongly
ionized.  This places it far from the Galactic center's intense UV
sources, and possibly outside the very center region itself.

All tracers in Fig.~\ref{fig:overview3labels}, but most obviously the
20\,cm image, show arcs and approximately circular structures with
diameters of a few tenths of a degree and approximately bisected in
longitude by the nonthermal filaments of the Radio Arc.  \cii\
integrated intensity in Fig.~\ref{fig:overview3labels}a clearly shows
the Arched Filaments, the set of arcs that lie above and approximately
parallel to the Galactic plane. \cii\ also follows an arc near the
dust emission ring surrounding the MSX Bubble.  All of the arcs are
also visible in 20\,cm radio continuum images (e.g.\ \citealt{lang10,
  heywood22}), indicating that at least some of the \cii\ is
associated with \hii\ regions.

Table~\ref{tab:Ncii} contains lower limits to \cpl\ and associated
proton (H$^+$ + H + 2\htwo) column densities and masses.  \cpl\
results are in the optically thin and thermalized limit for
$T_{ex} \gg 91$\,K \citep{crawford85, goldsmith12} and unit areal
filling factors for the specified regions.  These assumptions, along
with missing emission around zero velocity, make the \cpl\ column
densities in Table~\ref{tab:Ncii} lower limits.  While low, the
estimates are likely reasonable within a factor of a few: the \cpl\
column would be 1.8 times larger for emission split equally between a
PDR and an \hii\ region with $T_{PDR} = 150$\,K and
$n_{PDR} = 10^{3.5}$\persqcm\ and $T_e = 6000$\,K and
$n_e = 200$\persqcm.

In converting from \cii\ intensity to hydrogen mass we assumed an
elemental abundance ratio of X(C)/X(H) = $3 \times 10^{-4}$ and that
all atomic carbon is singly ionized. This is an intermediate value
that is likely accurate within a factor of two: \citet{garcia21}
derived an abundance ratio of $7 \times 10^{-4}$ from PDR modeling of
the Arches region, and \citet{sofia04} reported $1.6 \times 10^{-4}$
along translucent sight lines in the Galactic disk.

Comparison with molecular column densities from \ceio\ $J = 2-1$
\citep{apexco} emission yields an upper limit of \cii\ to molecular
proton column density ratio of 0.12.  This limit follows when all of
the \ceio\ emission is from the same region as \cii, and is optically
thin and in local thermodynamic equilibrium at an excitation
temperature of 150\,K, typical of the bulk material in PDRs. The ratio
decreases approximately linearly with temperature, to 0.03 for \ceio\
at the 35\,K temperature representative for Galactic center cloud
bodies. This column depth comparison is quantitative evidence that
\cii\ is from relatively thin layers on the surfaces of background
molecular clouds.  This \cii/molecular column density ratio scales
linearly with the assumed X($\mathrm{^{18}O)}$/X($\mathrm{^{16}O}$)
abundance ratio of 1/245 for the Galactic center \citep{wannier89} and
a representative X(CO)/X(\htwo) abundance ratio of $10^{-4}$.

The general \cii\ surface brightness across the \sgra\ region is
comparable to that of the \sgrb\ region \citep{harris21}.  In both
fields the extended emission accounts for nearly all of the total
emission.  The CND and +50\kms\ cloud are considerably brighter than
the extended emission, but their small sizes make them negligible
contributors to the total \cii\ flux. In our Galactic center, \cii\
reflects the amount of distributed radiation rather than identifying
the most intense sources of radiation, no matter how luminous.

\begin{deluxetable*}{rrrrrrr}
\tabletypesize{\scriptsize}

\tablecaption{Summary of areas, minimum \cii\ column densities, and
  implied proton column densities and masses in selected regions in
  the \sgra\ region.  The dashed line in Figure~\ref{fig:moments}a
  marks the $\ell = 0.17^\circ$ dividing line between the sub-regions
  in table lines 2 and 3. ``Arches region with \hii'' refers to the
  region containing the Arched Filaments and the bright \hii\ regions at the
  bases of the E and W filaments from the bright \hii\ regions
  (G\,0.09+0.01, G\,0.07+0.04, and G\,0.10+0.08, \citealt{pauls76});
  ``Arches region without \hii'' excludes the areas containing these
  bright \hii\ regions.  cgs intensity units are
  $\mathrm{erg~s^{-1}~cm^{-2}~sr^{-1}}$, multiply by $10^{-3}$ to get
  $\mathrm{W~m^{-2}~sr^{-1}}$.  Absorption near zero velocity,
  approximations of $T_{ex} \gg 91$\,K and optically thin emission
  make the \cii\ column densities and masses lower limits. $N$ and
  $M$(H$_{C^+}$) refer to the column density and mass of protons
  associated with the ionized carbon.  \label{tab:Ncii}}

\tablewidth{0pt}
\tablecolumns{7} 
\tablehead{
\colhead{Region} & 
\colhead{Area} &
\colhead{$I$(\cpl)} & 
\colhead{$N$(\cpl)} & 
\colhead{$N$(H$_{C^+}$)} &
\colhead{$M$(\cpl)} & 
\colhead{$M$(H$_{C^+}$)}  
\vspace{-3mm}\\
\colhead{} & 
\colhead{pc$^2$} & 
\colhead{$10^{-4}$\,cgs} &
\colhead{$10^{17}$\,cm$^{-2}$} & 
\colhead{$10^{21}$\,cm$^{-2}$} & 
\colhead{M$_\odot$} & 
\colhead{$10^{3}$\,M$_\odot$}
}
\startdata
Entire region & 3302.8 & 14.4 & 8.9 & 3.0 & 296.2 & 111.9 \\ 
$-0.15^\circ \leqslant \ell \leqslant 0.17^\circ$ & 1879.2 & 15.4 & 9.6 & 3.2 & 180.3 & 68.1 \\ 
$0.17^\circ \leqslant \ell \leqslant 0.41^\circ$ & 1423.6 & 13.1 & 8.1 & 2.7 & 115.9 & 43.8 \\ 
Arches region with \hii\ & 328.0 & 17.7 & 11.0 & 3.7 & 36.2 & 13.7 \\ 
Arches region without \hii\ & 233.5 & 14.7 & 9.1 & 3.0 & 21.4 & 8.1 \\ 
Circum-Nuclear Disk & 3.3 & 74.0 & 46.0 & 15.3 & 1.5 & 0.6 \\ 
+50 cloud & 2.5 & 32.2 & 20.0 & 6.7 & 0.5 & 0.2 \\
\enddata
\end{deluxetable*}

\begin{figure}[t]
\centering
\includegraphics[width=0.48\textwidth, trim=28 149 33 68, clip]{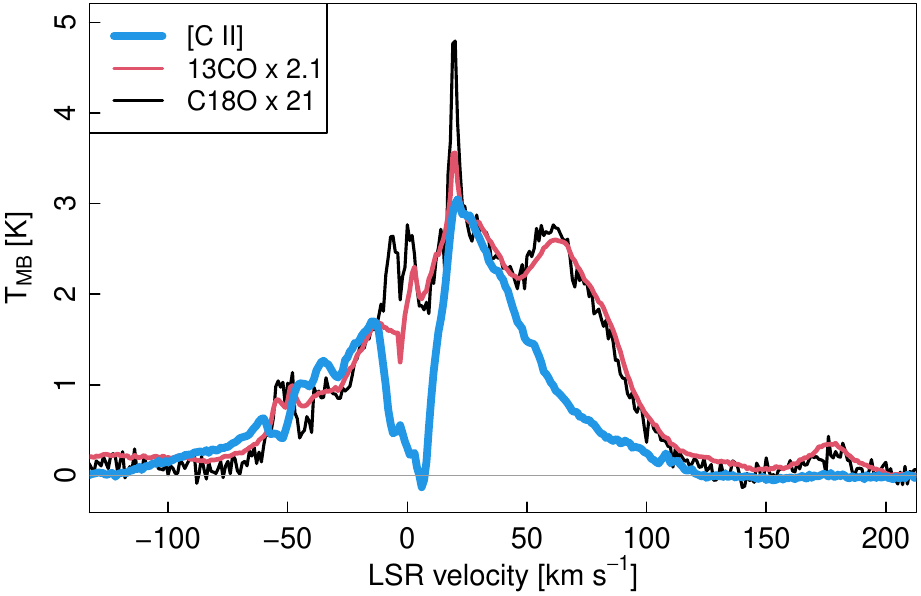}  
\caption{Region-averaged \cii\ spectrum at 1\kms\ velocity resolution,
  overlaid on \thco\ and \ceio\ $J = 2-1$ spectra for the same
  region. The lack of \cii\ emission near zero velocity is due to
  absorption along the line of sight through the Galactic plane. The
  slightly negative amplitude of the absorption is most likely due to
  baseline fitting errors, possibly with some contribution from
  emission in the distant reference position, which is much weaker
  than emission from the \sgra\ region itself. \label{fig:meanSpect}}
\end{figure}

\subsection{Velocity structures \label{ssec:velfield}}
Velocity information from the \cii\ line is essential for
distinguishing the Galactic center's different physical components
along the line of sight. Figure~\ref{fig:meanSpect} is the \cii\
spectrum averaged over the entire region, showing the wide velocity
extent characteristic of the Galactic center.  As toward \sgrb\
\citep{harris21}, the \cii\ and CO isotopologue lineshapes generally
agree except around zero velocity from \cii\ absorption along the line
of sight through the Galactic plane and from a low-ionization
molecular cloud near 55\kms\ lying between
$-0.13\degr \lesssim b \lesssim 0.01\degr$ and
$\ell \gtrsim 0.18\degr$.  Its lineshape and ionization are
similar to the +90\kms\ cloud toward \sgrb\ \citep{harris21}.  Other
than emission from the low-ionization cloud between 50 and 100\kms,
the differences in spatial distribution but similarities in velocity
structure between the CO isotopologue and \cii\ lines indicates that
their emission is from related regions, but with varying excitation
conditions apparent in different lines.

\begin{figure}[t]
\centering
\includegraphics[width=0.48\textwidth, trim=78 111 95 155, clip]{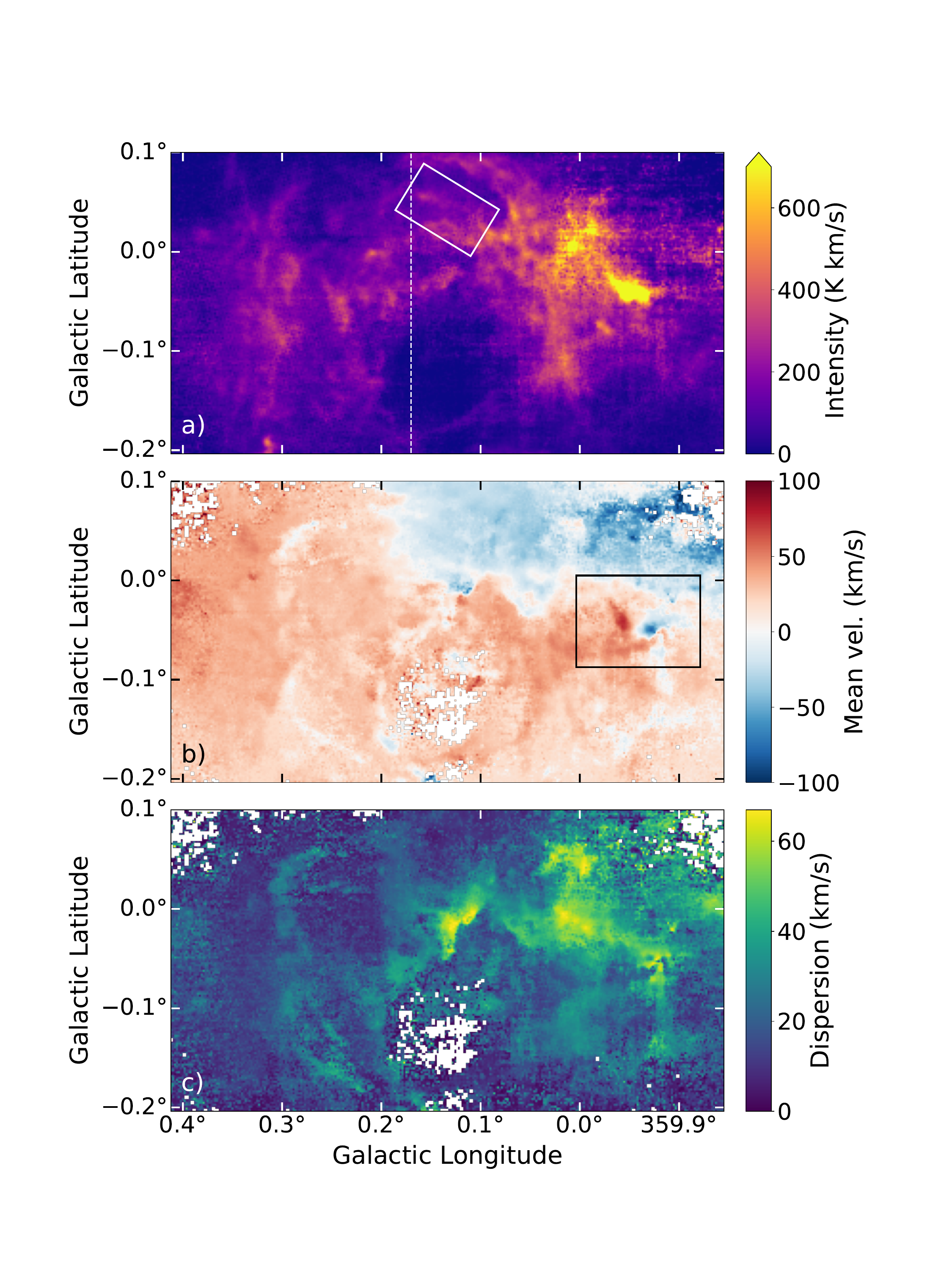}
\caption{\cii\ moment images on linear intensity scales.  Top:
  Integrated intensity (moment 0) over $\pm120$\kms. Middle:
  Intensity-weighted mean velocity (moment 1). Bottom:
  Intensity-weighted velocity dispersion (moment 2).  The blank areas
  near $\ell \approx 0.13\mdeg$ and in the upper corners of panels b)
  and c) are regions where the signal-to-noise ratio fell below the
  $3\sigma$ cutoff per velocity channel.  The white box in panel a) is
  the region (in equatorial coordinates) containing images of the E1
  and E2 Arched Filaments in multiple fine structure lines that we
  discuss in Sec.~\ref{sec:irfsl}.  The black box in panel b) shows
  the region around the CND that we discuss in Sec.~\ref{sec:cnd}.
  The dashed vertical line in a) is at $\ell = 0.17^\circ$.
  \label{fig:moments}}
\end{figure}

At finer scale, \sgra\ region's velocity field is complex, containing
multiple features across the field we imaged.
Figure~\ref{fig:moments} provides an overview of \sgra\ region's
velocity field with images of moments 0 through 2 across the field:
the velocity-integrated intensity, intensity-weighted mean velocity,
and intensity-weighted velocity dispersion.  The dominant velocity
component in panel b) has a characteristic velocity of +28\kms,
trending smoothly in the sense of Galactic rotation with extrema of
39\kms\ at the positive longitude edge of the image to 27\kms\ at the
negative longitude edge.  We call this extended background cloud the
``+28\kms cloud'' in this paper.  A second major component dominates
the upper right region of the field, with velocities peaking in the
$-20$ to $-40$\kms\ range.  Such velocities are characteristic of the
molecular cloud associated with the Arched Filaments.  As has been
noted previously (e.g., \citealt{serabyn87, lang01}), the latter
material has forbidden velocities in the sense of Galactic rotation.

The southwestern (equatorial coordinates; $+\ell$ in Galactic
coordinates) and northeastern (in equatorial coordinates; $-\ell$
in Galactic coordinates) lobes of the CND stand out in both intensity
and velocity at $(\ell, b) \approx (-0.04\mdeg, -0.05\mdeg)$.  The
middle (velocity) panel highlights extended emission at the lobes'
velocities that stretches away from both lobes, apparently diagonally
from the plane of the CND. Section~\ref{sec:cnd} contains a more
thorough discussion of emission near the CND.

Lighter regions in the bottom panel mainly show where the lines have
two distinct components. Lineshape comparison with \ceio\ and \thco\
$J = 2-1$ lineshapes \citep{apexco} confirms that these are truly
separate velocity components, and are not an interruption of a single
broad component by a line-of-sight absorption feature. A clear example
of a region with two components is the vertical stripe with curved
ends at $\ell \approx 0.28\mdeg$, which is matched with a lower
average velocity in panel b).  A similar pattern is present in panel
c)'s bright region shaped like a rearing unicorn centered on
$(\ell, b) \approx (0\mdeg, 0\mdeg)$, which marks an overlap region
between two velocity components. Correlations between moments 0 and 2
are particularly prominent from regions with two separate spectral
components of comparable intensity.  These add in intensity and the
velocity separation increases the total dispersion.

\begin{figure*}[!ht]
\centering
\includegraphics[width=0.95\textwidth, trim=43 79 61 96, clip]{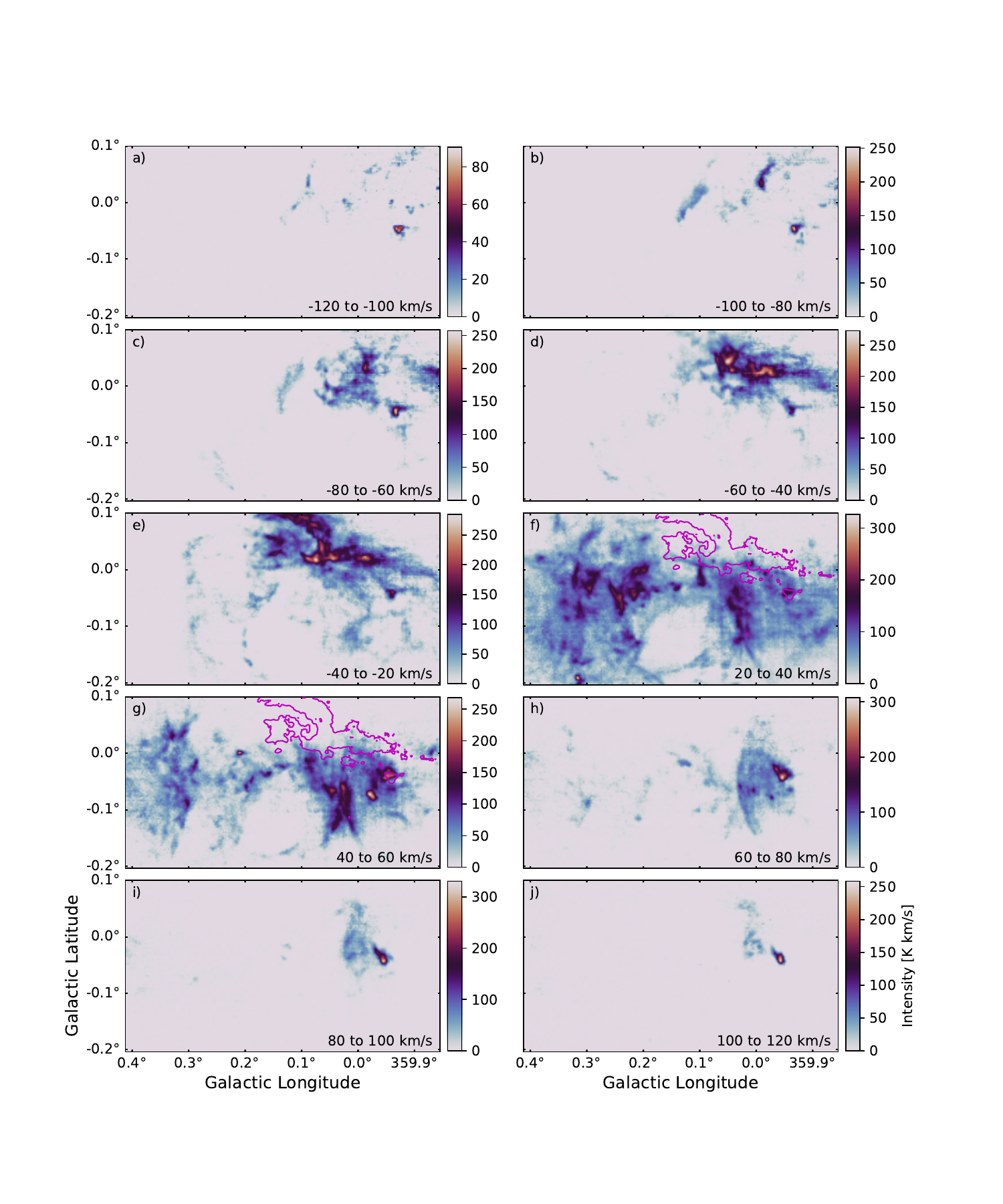}  
\caption{Array of integrated intensity (moment 0) channel images over
  20\kms\ velocity ranges. Integrated intensity scales vary by panel,
  as indicated by each scale bar.  The color palette helps show the
  extended emission.  The velocity range $-20$ to $+20$\kms\ is
  affected by Galactic plane absorption along the line of sight, and
  is not included in this set.  Magenta contours in panels f) and g)
  follow the outline of emission from the Arches cloud in panel e);
  see text.
  \label{fig:channels}}
\end{figure*}

Figure~\ref{fig:channels} is an array of \cii\ integrated intensity
channel maps covering successive 20\kms\ velocity ranges.  The western
lobe of the CND is the compact source at
$(\ell, b) \approx (-0.07\degr, -0.05\degr)$ in panels a) through c),
which cover $-120$ to $-60$\kms.  The lobe itself is brightest at
$-75$\kms, in panel c).  Wisps and areas of emission appear at a
variety of velocities across these panels, but with no obvious
connection to the CND's western lobe.  In panels b) and c) a small
curved cloud with velocity peaking at $-85$\kms\ stretches from
G\,0.09+0.01 at the base of the E Arched Filaments toward the base of
the Sickle handle, with center at
$(\ell, b) \sim (-0.1\degr, -0.0\degr)$.  This cloud is also notable
as the brighter region forming the unicorn's forelegs in
Fig.~\ref{fig:moments}c)'s moment 2 image.

Emission peaking from $-57$ to $-44$\kms, brightest toward the large
\hii\ region complex G\,$-0.07$+0.04 north of the \sgra\ radio
continuum peak, emerges in panel d).  A narrow absorption notch at
about $-55$\kms\ across the region indicates that this material is on
the far side of the bar-driven 3\,kpc arm \citep{sormani15, li22},
placing it within the Galactic center.

\begin{figure}[t]
\centering
\includegraphics[width=0.48\textwidth]{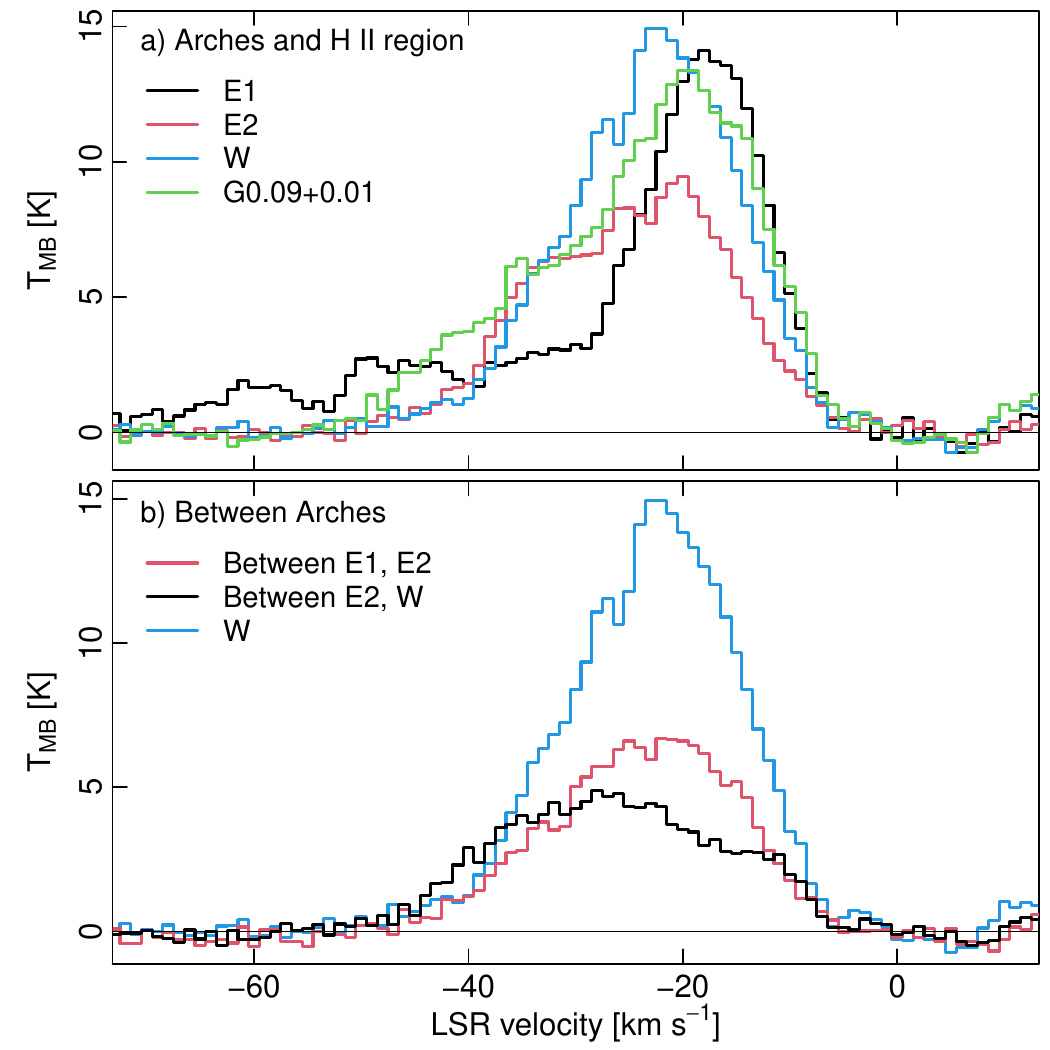} 
\caption{Top: Typical \cii\ spectra of the E1, E2, and W Arched
Filaments and the G\,0.09+0.01 \hii\ region at the base of the E
filaments.  Bottom: \cii\ spectra of typical regions between the E1
and E2 filaments, the E2 and W filaments, with the W filament spectrum
repeated as a reference.  These comparisons indicate that emission
from the Arched Filaments adds to more distributed \cii\ intensity
from the background Arches molecular cloud.
\label{fig:archesStack}}
\end{figure}

Emission from $-40$ to $-20$\kms, visible in panel e), is dominated by
the Arched Filaments and their background molecular cloud.
Figure~\ref{fig:archesStack} shows typical spectra from the bright
filaments comprising the E1, E2, and W Arched Filaments.  The regions
are comparable in peak brightness and velocity distributions with each
other and with the G\,0.09+0.01 \hii\ region at the base of the East
filament.  These spectra are typical for all of the bright emission in
the Arches region.  The lower panel in Fig.~\ref{fig:archesStack}
highlights the \cii\ intensity between the brighter filaments: The
Arched Filaments are brightness enhancements above a substantial
background.

Both \thco\ and \ceio\ $J = 2-1$ lines have two distinct velocity
components centered at $-35$ and $-13$\kms\ across the Arches region.
These components are also present in CS from the background cloud
(Peak 2 in \citealt{serabyn87}).  \cii\ lineshapes share the total
velocity range of these the components, usually peaking between them,
but sometimes appearing lumpier or flatter as one component or the
other dominates.  Bright \cii\ emission also shares the spatial
distribution and velocity parameters of H92$\alpha$ \citep{lang01},
indicating that \cii\ is associated with ionized plasma as well as
molecular gas.  We discuss the Arched Filaments in more detail in
Sec.~\ref{sec:circles}. 

\begin{figure}[h]
\centering
\includegraphics[width=0.47\textwidth, trim=28 149 33 68,
clip]{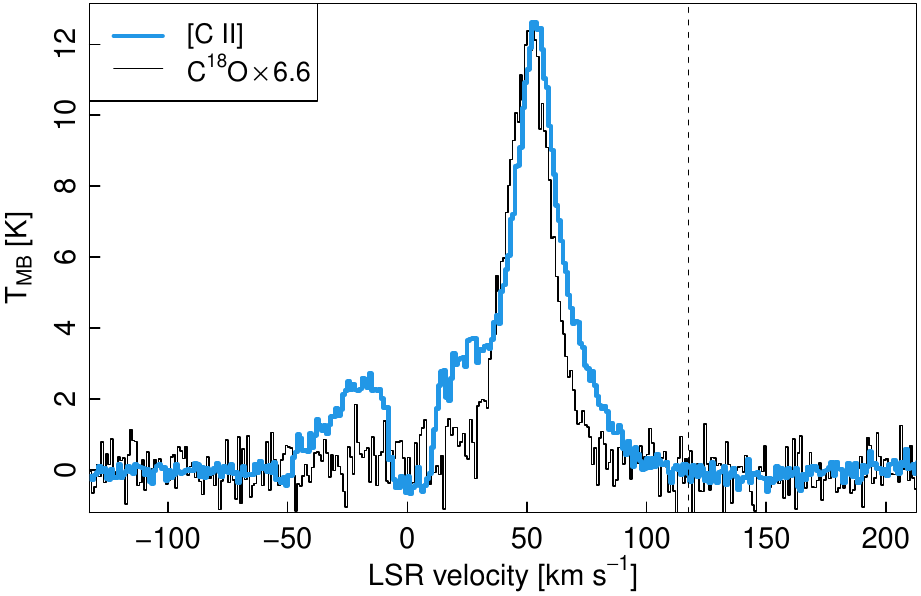}  
\caption{\cii\ and \ceio\ $J=2-1$ spectra of the bright \cii\ peak
  region of the +50\kms\ cloud over a 0.6 by 1.0 arcminute elliptical
  region at 30\degr\ position angle, centered at ($\ell, b$) =
  ($-0.025\degr, -0.076\degr$). The \cii\ line has a FWHM width of 28\kms\
   centered at 54\kms. The vertical line marks the position of the
  [$^{13}$C\,{\sc ii}] $F = 1-1$ hyperfine line component, which is
  not detected. \label{fig:p50}}
\end{figure}

Returning to the channel maps, and skipping the velocity range from
$-20$ to $+20$\kms\ because of absorption from the Galactic plane, the
series continues to show the most spatially extended emission, from 20
to 40\kms, in panel f).  The cloud velocity peaks run from 27\kms\
toward the \pell\ side of the panel to 39\kms\ at the \mell\ edge.  A
magenta contour in the panel outlines the extent of the negative
velocity emission from panel e), which appears to fit neatly above and
to the right of the brightest emission in panels f) and g).  A
geometrical match could be the result of absorption by dust of
background emission, a background cloud shielded from foreground UV,
or coincidence.  An absence of \cii, \thco, and \ceio\ $J=2-1$ at
velocities near 30\kms\ within the contour points to coincidence as
the most likely explanation.

The limbs of the MSX Bubble are clear at 70\um\ and in \cii\ at
velocities between 20 and 60\kms.  This velocity range is common for
much of the Galactic center emission, indicating that the MSX Bubble
is also in the center, and not somewhere in the Galactic plane.
Integration over the entire apparently empty region shows only weak
\cii\ emission spatially and spectrally associated with the +50\kms\
cloud, which extends through the Bubble's center \citep{fukui77}.
Without detection of velocities corresponding to a front or back, it
does not appear to be an expanding sphere within a larger cloud,
although it could be an expanding hole in a thin sheet extending
across the line of sight.

Panel g) shows 40--60\kms\ line wing emission lingering from panel f)
as well as growing emission from the CND. A cloud associated with the
Sickle is visible at $(\ell, b) \approx (0.18\mdeg, -0.05\mdeg)$ as
the leftmost of 3 emission knots bordering the Bubble. Its velocity
peak at 38\kms\ and linewidth of about 40\kms\ are characteristic of
the brighter \cii\ regions to more positive longitudes. Neither \cii\
intensities nor lineshapes give any hint of interactions involving
magnetic fields associated with any of the center's nonthermal
filaments, whether in and near the Sickle, in the regions where the
Arched Filaments seem to cross the Radio Arc's nonthermal filaments,
or anywhere else.  This result confirms and refines the conclusion by
\citet{poglitsch91} that there is no obvious sign of interaction
between the magnetic fields of the Radio Arc's nonthermal filaments
and \cii\ excitation.

The core of the +50\kms\ cloud (M--0.02--0.07; e.g., \citealt{fukui77,
  guesten81, mezger86}) is the second bright compact peak from the
right in panel g). This peak is close in position to CO $J=7-6$ peaks
in cross-cuts across the limb of \sgrae\ in \citet{genzel90}.
Figure~\ref{fig:p50} compares the spectra of \cii\ and \ceio\
$J = 2-1$ from the APEX telescope averaged over the brightest \cii\
emission contained in an elliptical area 1\amin\ long and 0.6\amin\
wide centered at ($\ell, b$) = ($-0.025\mdeg, -0.076\mdeg$).  Center
velocities and widths in the line core agree well, associating the
main \cii\ emission with the core of the +50\kms\ cloud. The \cii\
line's center velocity of 54\kms\ is slightly higher than the 50\kms\
reported for the ammonia core by \citet{guesten81} and the surrounding
region seen in \ceio\ $J = 2-1$ and CO $J=7-6$ reported by
\citet{genzel90}.

The spectrum covers the velocity of the [$^{13}$C\,{\sc ii}] $F = 1-1$
hyperfine satellite component, which is offset by +63.2\kms\ from the
[$^{12}$C\,{\sc ii}] line center \citep{cooksy86}.  Spectral binning
of 20\kms\ centered on the $^{13}$\cpl\ velocity gives an intensity
ratio $I([^{12}$C\,{\sc ii}])/$I([^{13}$C\,{\sc ii}]) $\geqslant 84$.
This ratio implies an optical depth in [$^{12}$C\,{\sc ii}] of
$\tau \leqslant 2.8$ for equal excitation temperatures for the two
lines, a Galactic center abundance ratio of X($^{12}$C)/X($^{13}$C) =
31 and 0.125 of the line intensity in the $F = 1-1$ component
\citep{wannier89, ossenkopf13}.  Much of the potential [$^{13}$C\,{\sc
  ii}] emission could be from the +50\kms\ cloud line wing, however,
for $\tau \leqslant 0.9$ if half or more of the intensity at the
[$^{13}$C\,{\sc ii}] velocity is from the wing.

\begin{figure}[h]
\centering
\includegraphics[width=.45\textwidth, trim=40 670 140 175,
clip]{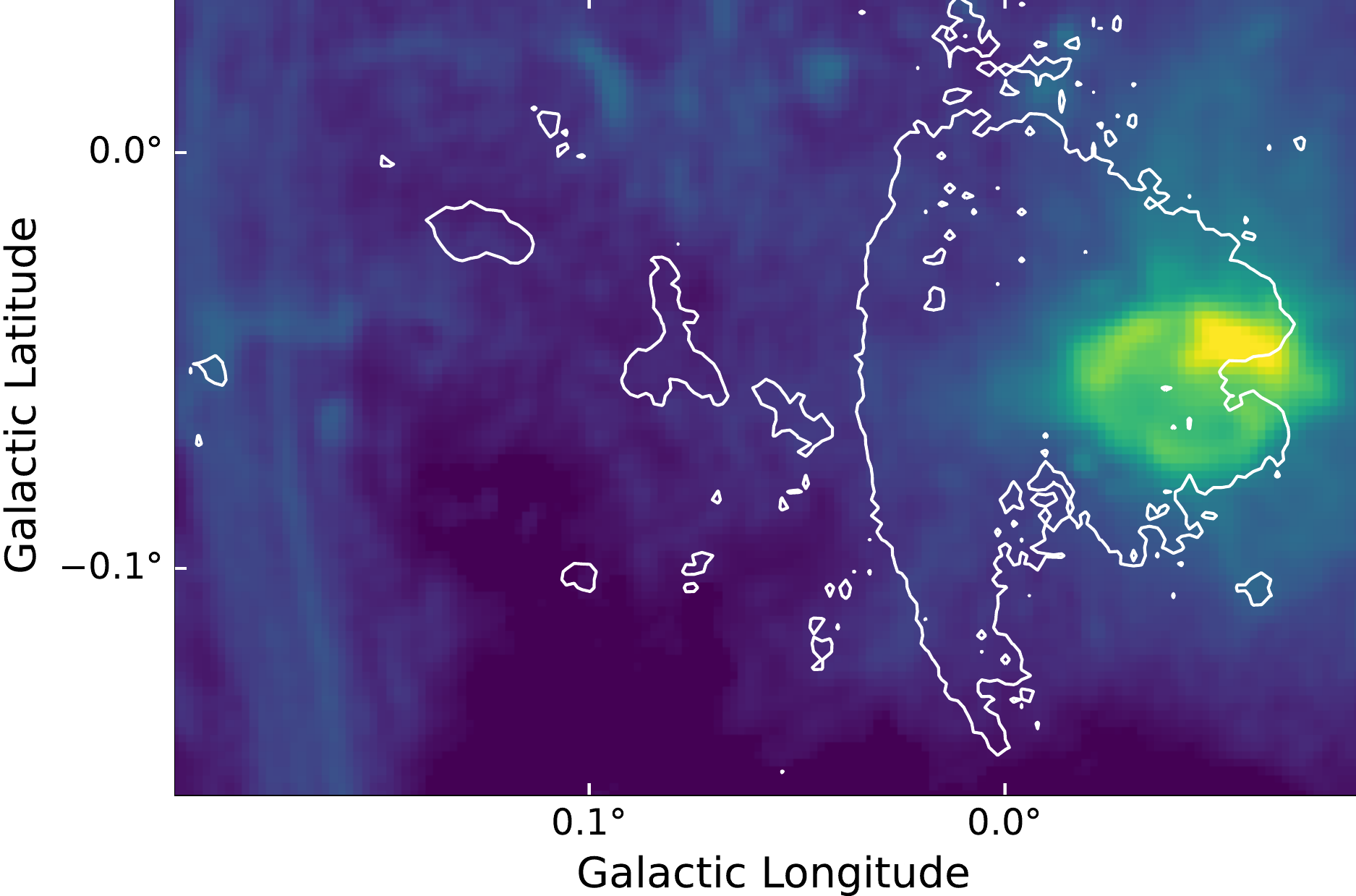} 
\caption{Extent of \cii\ emission from 60--80\kms\ from
  Fig.~\ref{fig:channels}g) superposed on 20\,cm radio continuum
  \citep{lang10} with a square-root color stretch to show the Radio
  Arc, \sgrae, and CND.  The curved edge of the \cii\ contour to
  $+\ell$ seems to be bisected by the faint tongue of 20\,cm emission
  that extends along \sgrae's major axis.
\label{fig:sgraEfan}}
\end{figure}

Continuing to 60--80\kms\ in panel h), a clearly-visible fan-shaped
structure some 530\asec\ long and 60\asec\ wide is reminiscent of a
bow shock or outflow.  This structure is visible in panel g) as well,
where it appears to touch a ridge of emission around the MSX Bubble
seen in \cii\ and 70\um.  Figure~\ref{fig:sgraEfan} superposes the
35\,K\kms\ contour delineating the main emission shown in
Fig.~\ref{fig:channels}'s panel h) on the 20\,cm radio continuum
image.  The contour wraps around the negative longitude side of
\sgrae, including the northeast lobe of the CND, then extends to
positive longitudes beyond the other side of \sgrae. The region within
the contour is filled with diffuse \cii\ emission at 71\kms\ peak
velocity. Additional emission along the lower half of the contour to
$+\ell$ is the wing of the velocity component that peaks at 44\kms\
visible in Fig.~\ref{fig:channels}g). This \cii\ emission is about
35\asec\ to $-\ell$ from the similar arc of IRAC 8\um\ emission at
$b \lesssim -0.5\degr$ \citep{stolovy06}.

The final panels i) and j), covering 80 to 120\kms, show the CND's western
lobe and high-velocity remnants of the fan-shaped emission in panel
h).  The CND lobe and the nearby teardrop-shaped diffuse emission near
$(\ell, b) \approx (0\mdeg, -0.03\mdeg)$ do not seem to be connected:
\cii\ from the western lobe is brightest at 75\kms, while the velocity of
the diffuse emission peaks at about 97\kms.  This high-velocity
component also appears as a brighter region in the moment 2 image of
Fig.~\ref{fig:moments}. There is no velocity bridge between the two
regions, and the velocities in the cloud around 100\kms\ have no
discernible systematic velocity gradient.

\subsection{Compact \hii\ regions}
In addition to \cii\ from large-scale distribution and \hii\ regions
in the Arches and \sgrae\ regions, we find two bright compact \cii\
sources with 20\,cm and 70\um\ counterparts.  The positions match
sources A and E of the \hii\ regions described by \citet{immer12}.
Other sources in Immer et al.'s list have no \cii\ counterparts,
although source D is bright at 20\,cm and 70\um.  Our \cii\ data
provide velocity information for sources A and E.

Immer et al.'s source A is visible just to positive $\ell$ from the
upper end of the Sickle at ($\ell, b) = (0.208\mdeg, -0.002\mdeg)$.
Spectrally, it appears as an 8\kms\ FWHM wide line centered at 46\kms\
on the positive velocity wing of the broader line from the surrounding
area, which peaks at about 30\kms.  Its peak brightness temperature is
20\,K.  Its diameter is about 25\asec, with some extension to $-\ell$.

A fuzzy region with diameter of about 50\asec\ at
($\ell, b) = (0.382\mdeg, 0.015\mdeg)$ contains the position of Immer
et al's source E. Its peak velocity is 35\kms, with width 15\kms, and
peak brightness temperature 9\,K. Its velocity is the same as that of
surrounding emission.

Velocities for sources A and E are consistent with locations in the
Galactic center, supporting Immer et al.'s conclusion that at least
source E contains massive star formation similar to Orion's M42 \hii\
region.

\section{Discussion}\label{sec:discuss}
\subsection{The Sgr\,A complex and [C\,II]  from the CND}\label{sec:cnd}
\begin{figure}[t]
\centering
\includegraphics[width=.45\textwidth, trim=175 66 127 121,
clip]{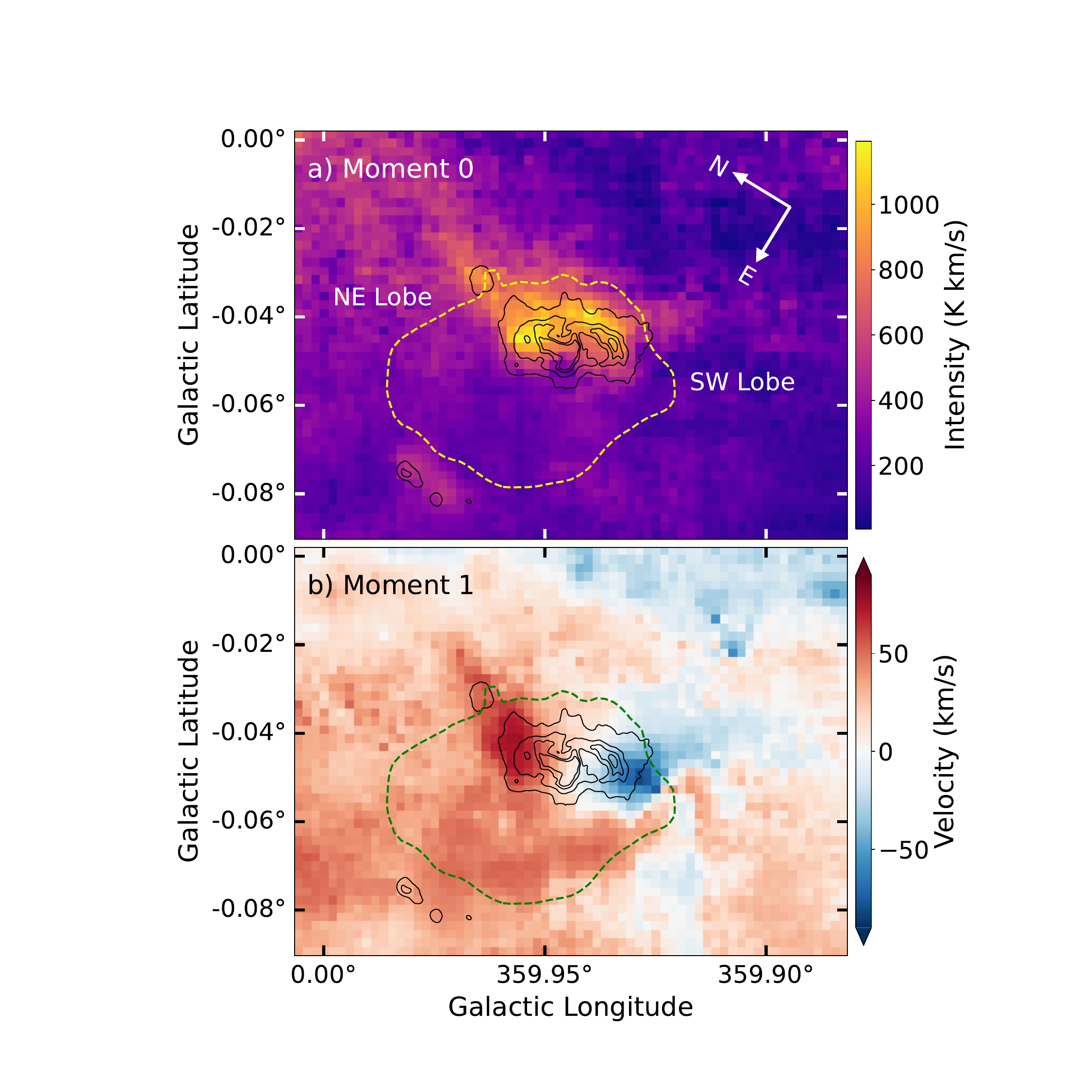} 
\caption{Views of the CND and nearby emission expanded from the boxed
  area of Fig.~\ref{fig:moments}b.  Top and bottom panels are the
  images of \cii\ integrated intensity (moment 0) and
  intensity-weighted \cii\ mean velocity (moment 1).  The \cii\
  extended emission stretching from the CND has a vee shape that seems
  to point toward the center of \sgrae, and also matches dust emission
  toward the northeast. In both panels solid contours show 37.1\um\
  emission that highlights the CND; the dashed contour line shows the
  outer edge of the \sgrae\ 20\,cm nonthermal radio continuum
  \citep{hankins20, lang10}.  Direction arrows and labels in panel a)
  indicate positions of the CND's edge-brightened limbs (northeast and
  southwest lobes) in equatorial coordinates.
  \label{fig:streamers}}
\end{figure}

Figure~\ref{fig:streamers} provides enlarged versions of the moments 0
and 1 images from the large-scale Fig.~\ref{fig:moments}b for a
detailed view of the area near the CND. Figure~\ref{fig:streamers}a
shows that, unlike molecular emission (e.g., \citealt{guesten87,
  christopher05, requenatorres12}), emission in both \cii\ and
37.1\um\ continuum (\citealt{hankins20}, shown in black contours) from
the CND's northeastern bright inner edge (lobe) is brighter than from
its southwestern lobe.  (We use equatorial coordinates in this
discussion of the CND to facilitate comparisons with the literature.
Fig.~\ref{fig:streamers}a shows the correspondence between the
coordinate systems.) Fainter \cii\ emission extends beyond the bright
inner edges of both the northeastern and southwestern lobes to radii
of about 6\,pc, in agreement with observations by \citet{iserlohe19}
and \citet{morrisCND}.  While \cii\ emission from the CND's inner
edges is brighter than that extending from the edges, the extended
emission covers larger areas, for integrated intensities about 1.6
times larger than at the inner edges of the CND (values in
Table~\ref{tab:Ncii}).

Both dust and plasma are present within the continuous \cii\ velocity
field extending from the edges of the CND. Emission to the northeast
follows the ridge of 37.1\um\ continuum, including the dust clump
about 0.02\degr\ (72\asec) to the equatorial north.  A deep 6\,cm
continuum image shows what \citet{zhao16} call a ``wing'' that shares
the spatial distribution of the northwestern \cii\ and dust emission.
Toward the southwest, however, there is little 37.1\um\ emission, and
the 6\,cm wing curves in the opposite direction to \cii.  Without
correspondence on both sides of the CND, it seems unlikely that the
\cii\ and 6\,cm wing emission trace the same physical conditions to
both sides of the CND.

Extended emission associated with the CND is more prominent in
velocity than in intensity.  Figure~\ref{fig:streamers}b is the \cii\
moment 1 (brightness-weighted mean velocity) image. Velocities at peak
brightness are about $+90$\kms\ at the CND's limb-brightened
northeastern lobe and $-75$\kms\ in the southwestern lobe, matching
intensity-weighted peak velocities in molecular emission toward the
CND (e.g., \citealt{harris85, guesten87, requenatorres12}), although
the \cii\ velocities do not decrease in the same way with radial
distance.  \grt\ \cii\ velocities are smaller than those inferred from
velocity-unresolved spectra of the CND by \citet{iserlohe19}, although
the spatial distributions generally agree.  The extreme \cii\
velocities of $\pm 120$\kms\ correspond to circular orbits near 2\,pc,
the approximate distance to the inner edge of the CND given the mass
distribution near \sgras\ \citep{genzel10}.  We compare the detailed
distributions of \oisw\ and \cii\ with other tracers in the CND in a
separate study by \citet{morrisCND}.
  
\begin{figure}[t]
\centering
\includegraphics[width=.45\textwidth]{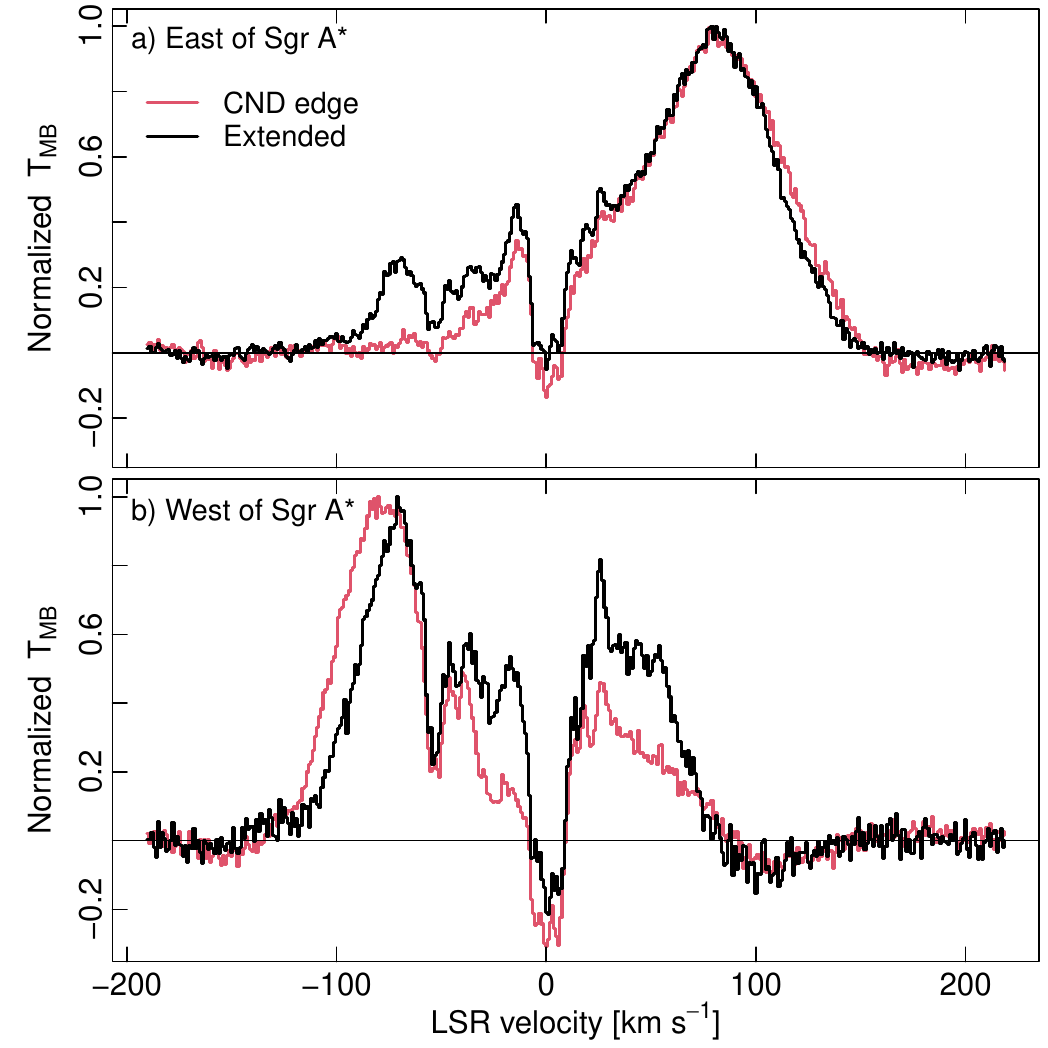} 
\caption{ \cii\ lineshape comparisons at the bright inner edges of the CND and
  the corresponding extensions to the east (+90\kms) and west
  ($-75$\kms). The small velocity shifts between the lobes and the
  corresponding extended emission complement the color images of
  Fig.~\ref{fig:streamers} that show very little velocity change with
  radius from the center of the CND.
  \label{fig:corewing}}
\end{figure}

Figure~\ref{fig:corewing} compares spectra integrated over the CND's
inner edges and related extended emission, showing no velocity change
with radius from the CND on the northeastern side, and little on the
southwestern.  If the material were in circular orbits in the mass
distribution around \sgras\ it would show velocity decreases of at
least 20\kms\ for the extreme velocities over the lengths of the
extended \cii\ emission, but this is not observed.  It is very
unlikely that the material is falling toward the center from larger
distances, as two independent streams would have to arrive at the CND
with velocities matching its rotating edges.  Instead, it seems likely
that the extended emission is material stripped from the CND.  

If the extended emission is driven from the CND by a wind from a
source within or outside the CND, the nearly constant velocity in the
extended emission places the driving source in a plane containing the
lines through the centers of the \cii\ emission and intersecting near
\ra{17}{45}{44.0}, \dec{-29}{00}{48}, and perpendicular to our line
of sight.  The lack of blueshift rules out an interaction with the
\sgrae\ supernova remnant, as had been suggested by \citet{zhao16}.
\citet{pedlar89} found that the CND appears in absorption against
nonthermal emission from \sgrae\ at 90\,cm, unambiguously placing the
CND in front of \sgrae: Any wind associated with sources in the
\sgrae\ nonthermal shell that accelerated material from the CND would
produce negative velocities compared with the CND.  It is possible
that the source of a wind stripping material from the CND is
associated with the exciting sources of the 7\amin-diameter \sgra\
halo, however: \citet{pedlar89} placed at least some of that halo in
front of the CND.

If the extended emission is stripped from the CND by a source outside
the CND it is not obvious why it has no signs of orbital motion about
the mass concentration centered on \sgras.  Material stripped in this
way would carry angular momentum and would stream purely radially.  If
a wind is responsible, a velocity well above the region's typical
100\kms\ orbital velocity would seem to be necessary to create and
perhaps maintain the linear features.

More likely, material at the inner edges of the CND is entrained by
the winds from massive stars in the Central Nuclear Cluster, and is
brightest where gas and dust densities are highest near the CND and
limb brightening is strongest. Outward radial flows into the in
inhomogeneous ISM near the CND would explain why \cii\ shows no
orbital curvature, and also the association with dust traced by
37.1\um\ continuum.  The dust distribution, especially the bright knot
to the northeast, is unlikely to be a consequence of acceleration by
winds.

\subsection{Excitation and origin of the Thermal Arched
  Filaments}\label{sec:circles}

The origin of the long and apparently coherent curved filaments that
make up the Arched Filaments has been a puzzle since their discovery
(e.g., \citealt{pauls76, yusefzadeh84, lang01, lang02}).
Figures~\ref{fig:channels}e and \ref{fig:archesStack} show that the
Arched Filaments are enhanced brightness regions (by a factor
$\sim 2$) on a larger background cloud.  While they stand out in dust
continuum emission at wavelengths shorter than 100\um, atomic lines,
and radio continuum (especially with the spatial filtering provided by
interferometers), the Arched Filaments are insignificant column
density peaks as traced by molecular emission (e.g.,
\citealt{serabyn87, genzel90, lang01, apexco}).  \cii\ flux from the
filaments is small, associated with only a few percent of the total
\cii\ flux across the \sgra\ complex.  The Arched Filaments appear to
be excitation peaks rather than physical structures.

\subsubsection{IR fine structure line excitation patterns \label{sec:irfsl}} 
\begin{figure}[t]
\centering
\includegraphics[width=0.48\textwidth, trim=385 75 125 105, clip]{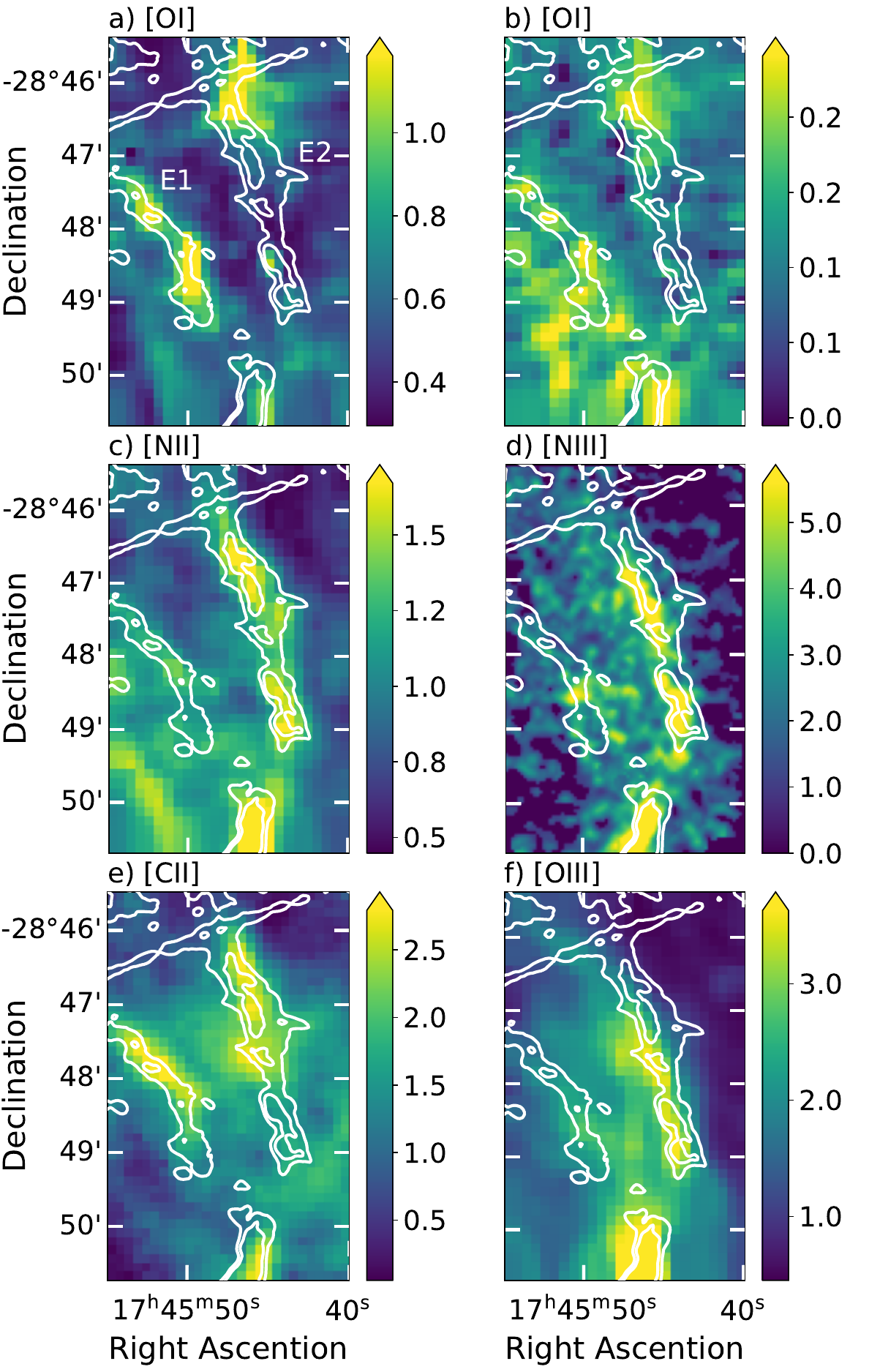}
\caption{ Far-infrared fine structure line intensity images of the E1
  and E2 Arched Filaments within the boxed region of
  Fig.~\ref{fig:moments}a. These images are in equatorial rather than
  Galactic coordinates.  Panels show: a) \oisw, b) \oilw, c) \nii\
  122\um, d) \niii\ 57\um, e) \cii\ 158\um, and f) \oiii\ 88\um\
  integrated line intensities.  These show differences in excitation
  between and along the filaments. The bright emission at the bottom
  of each image is the upper tip of the bright \hii\ region
  G\,0.095+0.012 at the base of E1 and E2. White contours are 20\,cm
  continuum from \citet{lang10}.  Transitions with different
  excitation conditions highlight different regions: E1 has
  systematically lower excitation than E2. Intensity scale bars are in
  units of $10^{-6}\, \mathrm{W~m^{-2}~sr^{-1}}$. Further information
  on each transition is in Table~\ref{tab:arches}.
  \label{fig:arches}}
\end{figure}

Observations of infrared fine structure lines show large- and
small-scale spatial changes in excitation conditions along and among
the Arched Filaments.  Figure~\ref{fig:arches} compares six infrared
fine structure line intensities from the E1 and E2 filaments.
Table~\ref{tab:arches} provides species, excitation, and other
information for each panel.  White 20\,cm continuum intensity contour
lines in each panel assist in spatial comparisons.

The run of excitation necessary for each line with location shows that
the E1 filament is dominantly neutral, while the E2 filament samples a
harder radiation field.  Peaks of \nii\ and \cii\ generally line up
well with 20\,cm radio continuum, although the two species only
partially overlap spatially.  Filament E1 and the northern end of E2
are brighter in \oi\ and \cii\ transitions that can trace neutral
material, but are weak in transitions of \nii, \niii, and \oiii\ that
require excitation energies above 13.6\,eV.  In contrast, the main
ridge of E2 is brighter in \niii\ and \oiii, both of which also tend
to peak approximately 20\asec\ toward the ``inside'' edge of E2 from
the 20\,cm, \nii, and \cii\ ridge line.  \niii\ is more extended than
\oiii\ along the ridge.  \oi\ and \oiii\ spatial distributions are
anticorrelated in E1 and E2, suggesting that spatially varying
ionization mechanisms convert \oi\ to \oiii.  The lack of spatial
overlap between the species suggests that they are in a relatively
thin layer, quite possibly on a surface.

Far-infrared fine structure line ratios provide no straightforward
discrimination between ionization by UV from stars or that produced by
shock waves.  Both photodissociation and shock models (e.g.,
\citealt{kaufman99, hollenbach89}) predict I(\oisw)/I(\ciiw) intensity
ratios greater than unity, but Fig.~\ref{fig:arches} shows ratios less
than unity.  This likely indicates that \oi\ is optically thick and
self absorbed, so considering its intensity alone is
incomplete. \oilw\ should be optically thin, however.  If the emission
were from a single PDR, the I(\cii)/I(\oilw) ratio of about 10
constrains the radiation field within about
$\mathrm{G_0} \approx 10^2-10^3$ Habing and particle densities of
about $10^4-10^2$\percucm\ \citep{kaufman99}. Reducing either the UV
field intensity or the particle density by an order of magnitude in
the PDR model would result in an intensity ratio of about 30, which is
representative of locations with weak \oilw\ emission.  The change in
ratio would more likely be due to a change in density than UV flux
since the radio continuum flux densities, which are linearly
proportional to Lyman continuum flux, are similar for all of the
Arched Filaments.  Our direct imaging is consistent with the
conclusion by \citet{garcia21} that the Arches region's \cii\ emission
is a spatially varying mixture of PDRs and \hii\ regions, with the
fraction of \cii\ from PDRs ranging from 25\% to 75\%.

\begin{deluxetable}{llrrrl}[t]
\tabletypesize{\scriptsize}
\tablecaption{Parameters for the spectral lines in the panels of
  Fig.~\ref{fig:arches}. $\lambda$ is the rest wavelength, EP the
  excitation potential in eV, T$_{\rm upper}$ the energy of the
  transition's upper level in K, and Source is the
  instrument that obtained the data. 
  \label{tab:arches}}
  \tablewidth{0pt}
  \tablecolumns{9}
  \tablehead{
    \colhead{Panel} &
    \colhead{Species and} & 
    \colhead{$\lambda$} &
    \colhead{EP} &
    \colhead{T$_{\rm upper}$} &
    \colhead{Source} \vspace{-3mm} \\ 
  \colhead{ } &
  \colhead{transition} &
  \colhead{\um} &
  \colhead{eV} &
  \colhead{K} &
  \colhead{} }
\startdata
a) & \oi~$\mathrm{^3P_{1}} - \mathrm{^3P_{2}}$ & 63.2 & 0 & 99 & PACS \\
b) & \oi~$\mathrm{^3P_{0}}- \mathrm{^3P_{1}}$ & 145.5 & 0 & 327 & PACS \\
c) & \nii~$\mathrm{^3P_{2}} - \mathrm{^3P_{1}}$ & 121.9 & 14.5 & 189 & PACS \\
d) & \niii~$\mathrm{^2P_{3/2}} - \mathrm{^2P_{1/2}}$ & 57.3 & 29.6 & 251 & FIFI-LS \\
e) & \cii~$\mathrm{^2P_{3/2}} - \mathrm{^2P_{1/2}}$ & 157.7 & 11.3 & 91 & GREAT \\
f) & \oiii~$\mathrm{^3P_{1}} - \mathrm{^3P_{0}}$ & 88.4 & 35.1 & 163 & PACS 
\enddata
\end{deluxetable}

Even with the \oiii\ emission peaking some 20\asec\ inside the E2
filament's \nii\ peak, the lines' intensity ratio yields an
approximate constraint on the radiation field if the emission were
from a \hii\ region.  Without any position corrections to account for
the shift between peaks, the \oiii/\nii\ intensity ratio has upper
limit of 3.6 along the \oiii\ E2 filament and the G\,0.095+0.012 \hii\
region at the base of the Arched Filaments.  The ratio is closely
unity in the diffuse emission at the edges of the field.  The
similarity of the peak ratio in the filaments and \hii\ region
suggests that the ionized portions of the Arched Filaments see
radiation fields characteristic of Galactic center \hii\ regions.
\hii\ region models from \citet{rubin85} as summarized in
\citet{ferkinhoff11} indicate that the ratios correspond to stellar
effective temperatures $T_{\rm eff}$ from an upper limit of about
35,800\,K in the bright regions to 34,800\,K in the diffuse material,
if the excitation were purely stellar.  Such temperatures would be
typical for a mid-O stellar spectrum.

The distribution of emission from species with different ionization
energies, coupled with the relatively low contrast between the
filament peaks and valleys in the images and
Fig.~\ref{fig:archesStack}, indicates small-scale changes in
excitation conditions on top of a smoothly varying base, pointing to
local as well as large scale excitation variations.  This is difficult
to reconcile with solely radiative excitation by a distant source,
which would vary only slowly over the area.  As we discuss next, while
the region contains some mixture of PDRs and \hii\ regions, shocks may be
important as well.

\begin{figure*}[!ht]
\centering
\includegraphics[width=0.99\textwidth, trim=68 388 101 419,clip]{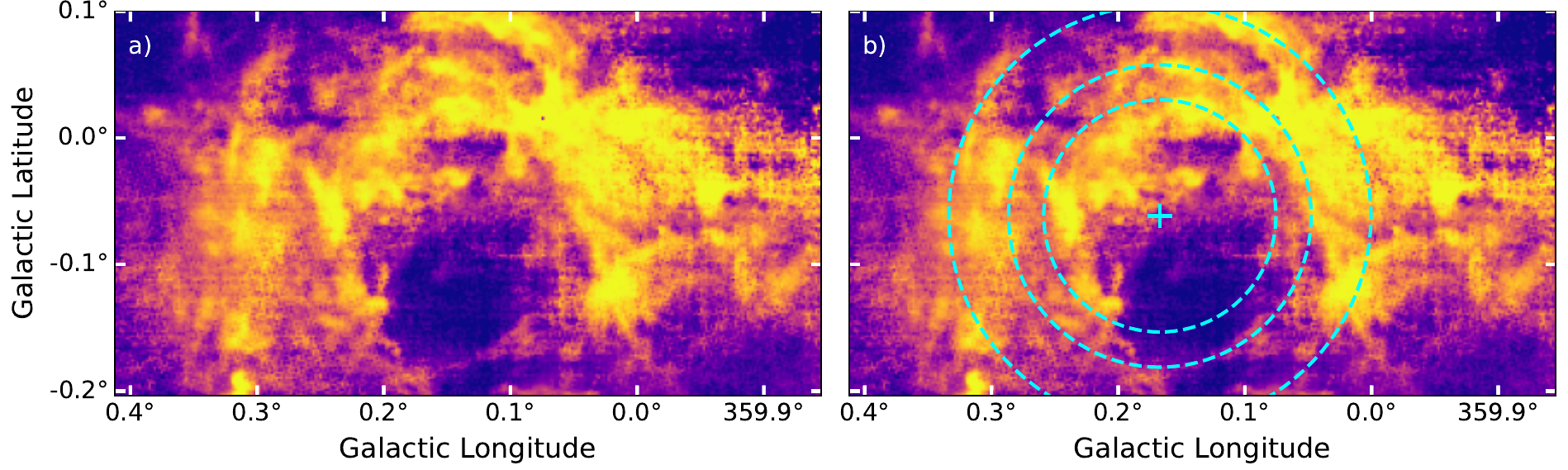}
\caption{Left: \cii\ integrated intensity from $-40$ to $+60$\kms\
  with histogram equalization to enhance contrast.  Right: The same
  image with superimposed circles with radii 330\asec, 430\asec, and
  600\asec\ (13, 17, and 24\,pc) centered on the Quintuplet cluster
  (marked by a cross) as a nominal center, and aligned with the E1,
  E2, and W Arched Filaments.  \label{fig:circles}}
\end{figure*}

\subsubsection{A common source produces the Arched Filaments' structure?}
The current main model for the Arched Filaments's structure is that
the ridges are protruding edges of a collection of clouds illuminated
by a distant source of UV \citep{lang01, lang02}.  This model for the
Arches region assumes a preexisting structure for the clouds, without
explanation for why the cloud edge distribution exists to produce
Arched Filaments with lengths of 10 to 20\,pc that are curved and
approximately concentric.  Dust polarization tracing the magnetic
field shows the same large-scale coherence and curvature as the Arched
Filaments \citep{morris92, chuss03, pare24}.  A formation mechanism
for the Arched Filaments involving tidal shearing is unlikely because
the background Arches molecular cloud does not share the curved shape
\citep{serabyn87}, and the Arched Filaments' center of curvature lies
some 20\,pc from the region's mass concentration in the Central
Nuclear Cluster.

At larger scale, single-dish radio continuum images of the \sgra\
region (e.g., \citealt{pauls76}) clearly show a circular region of
nonthermal emission some 20 to 30\,pc in radius centered on the area
containing the Sickle, Pistol, and Quintuplet cluster at about
$(\ell, b) = (0.15, -0.05)$.  Emission from the Arched Filaments wraps
around the edge of this \halo.  Fig.~\ref{fig:overview3labels}c) shows
some of the extended emission from the \halo, although interferometers
resolve out extended flux to reduce its prominence while better
showing the individual filaments (e.g., \citealt{yzmc84, lang01,
  heywood19}). \cii\ emission shares the approximately bull's-eye
pattern of multiple arcs with radio and dust continuum counterparts
that includes the Arched Filaments.

Even accounting for the human brain's predisposition for inventing
patterns where none exist, the concentric symmetry of the \sgra\
region appears to reflect physical reality.  Motivated by the Galactic
center's dynamic nature, we investigate whether energy from
centrally-concentrated transient sources could cause the shape of the
Arched filaments.
  
\begin{figure}[t]
\centering
\includegraphics[width=0.47\textwidth, trim=28 149 33 68,
clip]{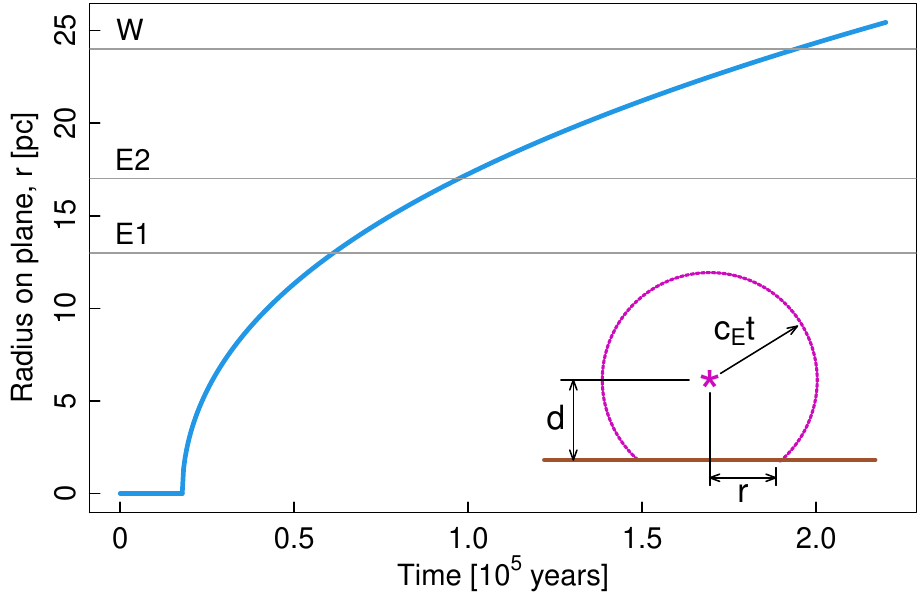}  
\caption{Radius $r$ vs.\ time of the circle on the surface of a plane
  cutting a Sedov-Taylor phase shock's spherical shell centered at
  distance $d$ from the plane, from eq.~(\ref{eq:roft}).  The inset
  figure shows the side-view geometry of this simple model.  The
  parameters for this particular curve are in the text.  Horizontal
  lines indicate distances $r$ to the Arched
  Filaments.  \label{fig:roft}}
\end{figure}

To explore the potential influence of occasional bursts of energy more
quantitatively, we consider a simple model of a point source emitting
a thin spherical shell of excitation that strikes a planar surface
representing the surface of a molecular cloud. Figure~\ref{fig:roft}
contains a side view of the geometry.  In this model many arcs,
including the Arched Filaments, are UV enhancements generated when the
excitation shell, whether a pulse of radiation or a mechanical shock,
strikes a molecular cloud surface and adds to the general UV
background from clusters of hot stars. Such a combination of local and
distributed excitation is consistent with the emission seen toward,
between, and around the Arched Filaments (e.g.,
Figs.~\ref{fig:channels} and \ref{fig:archesStack}). With relatively
static geometry, arcs tracing an excitation shell will have a common
center of curvature but different radii depending on the distance
between the central source and cloud surface.  Arc lengths will be set
by the surface's extent across the line of sight.  \cii\ and other
UV-excited emission will closely retain the radial velocity of the
cloud.  We discuss the strengths and weaknesses of this simple model
in Sec.~\ref{sec:procon} after considering the model itself more
thoroughly.

To better separate the Arched Filaments and other arcs from the
region's diffuse background, Figure~\ref{fig:circles} shows histogram
equalized \cii\ images with enhanced contrast.  Minimizing distraction
from extended emission and including a velocity range that covers
multiple cloud surfaces makes the circular patterns more obvious to
the eye in Fig.~\ref{fig:circles} than in the channel maps of
Fig.~\ref{fig:channels}.  All panels in Fig.~\ref{fig:overview3labels}
show this circular pattern to at least some degree as well. While a
general approach would be to search for segments of ellipses, segments
of circles are good representations of the Arched Filaments'
structure.  Fitting circles by eye to the various \cii\ arcs yielded
pattern centers in the vicinity of the Quintuplet cluster.  Circles in
the right-hand panel of Fig.~\ref{fig:circles} use the Quintuplet
cluster, position marked by a cross, as a convenient nominal center.
We discuss the role the cluster itself may play in
Sec.~\ref{sec:procon}. Circles with radii of 330\asec, 430\asec, and
600\asec\ (13, 17, and 24\,pc at 8.2\,kpc) lie on the E1, E2, and W
Arched Filaments (labeled in Fig.~\ref{fig:overview3labels}).  A
curved emission ridge toward the bottom center of the image lies from
about 4 to 6 o'clock on the 430\asec\ circle.  This ridge could be a
boundary to the MSX Bubble, although another curved dust ridge at
slightly larger Bubble radius better matches the overall curvature of
the MSX Bubble.  Toward positive longitude, the brightest \cii\
emission falls between the 430\asec\ and 600\asec\ circles. A spur of
emission at about 10 o'clock near the 600\asec\ circle may also be
part of a circular arc.

Following changes in structure over time constrains the speed of
propagation of the excitation shell in this simple model.
Figure~\ref{fig:roft} sketches the model and time evolution of the
circle with radius $r$ cut by a plane at distance $d$ from the center
of a thin spherical shell ballooning radially at expansion speed
$c_E$.  From the geometry in the figure's insert,
\begin{equation}
  r = \sqrt{\left( c_E \, t \right)^2 - d^2} \;, 
\label{eq:roft}
\end{equation}
with the circle expanding across the surface of the plane at speed
\begin{equation}
  \frac{\mdiff r}{\mdiff t} =
  \left(c_E + \frac{\mdiff c_E}{\mdiff t} t \right) \;
  \sqrt{ 1 + \left( \frac{d}{r} \right)^2 } \;
\label{eq:circle}
\end{equation}
for constant distance $d$.  

Equation~(\ref{eq:circle}) shows that the sphere's expansion speed is
a lower limit to the circle's expansion speed.  At large $d/r$ the
shell's wave front is almost flat when it strikes the plane, sweeping
over the region bounded by $r$ nearly instantaneously.  The velocity
in eq.~(\ref{eq:circle}) is similar to a phase velocity, with no
violation of causality even if the sphere's expansion speed $c_E$ is
the speed of light because different parts of the circle cannot
communicate with each other faster than the expansion speed $c_E$.

The circular, rather than elliptical, patterns we observe imply that
we view the model's plane at nearly normal incidence.  Whatever the
model, however, 20\,cm continuum images taken 17 years apart rule out
formation of the arcs as light echoes.  Between \citet{lang10}'s
observations in 2001 and \citet{heywood22}'s 2018 observations, light
would travel at least 131\asec\ across the line of sight for a
Galactic center distance of 8.2\,kpc.  Emission peak positions along
the Arched Filaments in the two images are coincident well within
5\asec, however, setting an upper limit for an expansion of a percent
or so of the speed of light.

Blast waves from supernovae fall within the appropriate speed range.
Such events are a possibility for producing occasional periods of
heightened excitation, with the Arched Filaments recording the impact
of blast waves on the surface of the Arches molecular cloud.  The
Galactic center has many of signs of supernova activity: most notably
the existence of the nonthermal \halo, the supernova remnant \sgrae,
an enhanced flux of cosmic rays, wind-blown lobes extending from the
center, and many other patterns of radio continuum circles and arcs
that are characteristic of supernova-driven bubbles throughout the
region.

For a supernova blast wave in its Sedov-Taylor phase,  
\cite{draine11} provides the shock velocity $v_s$  as
\begin{equation}
  v_s = 1950 \, \mathrm{km\,s^{-1}} \; E_{51}^{1/5} \;
  n_0^{-1/5} \;  t_3^{-3/5}  \;,
  \label{eq:vsedov}
\end{equation}
where $E_{51}$ is the energy provided by the supernova in units of
$10^{51}$\,erg, $n_0$ in \percucm\ is the particle density of the
medium the shock expands into, and $t_3$ the time in thousands of
years.  Solving eq.~(\ref{eq:roft}) for time and inserting
eq.~(\ref{eq:vsedov}) for $c_E$, the time required to reach radius $r$
on the plane is
\begin{equation}
  t_{r}(r) = 178 \, \mathrm{years} \; \left( r^2 + d^2 \right)^{5/4}
  \left( \frac{n_0}{E_{51}} \right)^{1/2} \;,
\label{eq:tr}
\end{equation}
with $r$ and $d$ in pc.  

Obtaining timing constraints from eq.~(\ref{eq:tr}) requires estimates
of the supernova energy, the particle density in the region, and the
actual distance $d$ from the exciting source to the Arches cloud.  We
take $E_{51} = 1$ as the nominal energy release of a 
supernova.  The average density in the central region is likely low:
\citet{mauerhan10} see no evidence for an intercloud interstellar
medium in the outflow of an LBV star in the Pistol region, and
measurements of H$_3^+$ by \citet{oka19} indicate low densities along
multiple lines of sight across the CMZ.  In the case we examine here,
supernovae would be frequent, so supernovae previous to those we
examine would likely have cleared much of the diffuse ISM across the
region.  Large-scale X-ray images show lobes characteristic of a
high-speed wind emerging from the Galactic center (e.g.,
\citealt{ponti19, predehl20, wang21, yeung24}), further evidence for
wind-swept low densities.  We take $n_0 = 0.1$\percucm\ along the path
toward the Arches cloud as a trial value, noting that propagation
times do not depend strongly on the exact $n_o/E_{51}$ ratio.

The gradual UV gradients along the Arched Filaments and across the
region imply that the sources of UV radiation produced by the massive
Arches and Quintuplet clusters must be some distance from the Arched
Filaments (e.g., \citealt{erickson91, rodriguez01, lang01, lang02,
  cotera05, hankins17}). In the absence of more precise information,
we take the distance $d$ between the supernova and the Arched
Filaments to be comparable to the Arches region's characteristic
dimensions, $d = 10$\,pc.

With these choices, shock fronts from three supernova events would
reach radii $r = 13$, 17, and 24\,pc on the plane in
$t_r = 6.1\times 10^4$, $9.7\times 10^4$, and $1.9\times 10^5$ years,
respectively.  Figure~\ref{fig:roft} shows this in graphic form: a
delay before the shock reaches the surface, then a sharp rise in
radius that bends to meet the circle's asymptotic expansion as
$r \gg d$.  The time between successive events would be
$3.6\times 10^4$ and $9.7\times 10^4$ years.  All of these values are
representative, and could easily vary by a factor of at least a few
depending on assumptions affecting the wind velocity and geometries.


\subsubsection{Strengths and weaknesses of the model \label{sec:procon}}

The simple model we have investigated is consistent with the overall
circular structures across the \sgra\ region and matches the
characteristics of other supernova interactions with molecular clouds.
It explains the curvature of the Arched Filaments and other arcs
without requiring cloud geometries that somehow produce parallel
curvatures over scales of many parsecs.  The varying excitation of
far-IR fine structure lines and the presence and positions of
high-excitation \oiii\ and \niii\ lines along the Arched Filaments is
easier to explain with shocks than with a uniform UV field from hot
stars.  Compression of magnetic field lines by supernova blast waves
can also explain the far-IR polarization pattern (as in e.g.,
\citealt{chuss03}) across the Arches region.

The Arched Filaments share similarities with IC\,443, a well-studied
interaction between a supernova blast wave and a molecular cloud.
\citet{castelletti11} measured IC\,443's spectral index between 74 and
330\,MHz to find that the flattest spectral indices,
$ -0.25 \lesssim \alpha \lesssim -0.05$ (for
$S_\nu\propto \nu^\alpha$), are in the interaction zone with the wind
at the edge of IC\,443 where the infrared lines are strongest.  The
high-resolution MeerKAT image of the spectral index across the region
\citep{heywood22} shows that $\alpha$ varies smoothly from about 0 in
the \hii\ regions at the bases of the Arched Filaments, to about
$-0.05$ to $-0.08$ along the bright ridges of the Arched Filaments and
other arcs traced by fine structure lines, to $-0.4$ and lower between
the filaments.  The western filament has a spectral index gradient
that is flatter toward its outer edge, and has less spectral index
variation in than the eastern filaments.

Shocks from fast winds striking molecular clouds produce UV that
excites mid- and far-infrared fine structure line emission
\citep{hollenbach89}.  Observations of supernova interactions with
molecular clouds, including IC\,443 (e.g.\ \citealt{burton90, reach00,
  neufeld07, millard21}), show enhanced emission across the interaction
zones, although observed intensities are sometimes difficult to
reconcile with theoretical models.  Emission from IR fine structure
lines from the Arches region is consistent with this structure.
Additionally, \citet{simpson1819} detected \oiv\ emission with {\it
  Spitzer} from clouds across the entire \sgra\ region, indicating
widespread energetic wind-driven shocks that produce higher energy
excitation than stellar UV alone can provide.

UV produced by shocks also heats dust.  In IC\,443, warm dust emission
peaks in the interaction region traced in shock-excited \htwo\
ro-vibrational lines \citep{vandishoeck93, burton88}.  Shocks shape
dust grains, with blast waves shattering and sputtering larger grains.
Dust also forms in post-shock ejecta (e.g.,
\citealt{lau15}). \citet{hankins17} found strong evidence for small
grains in their study of dust emission from the Arches region: To
match the flux and thermal structure revealed by their mid-IR images
of the Arches, their modeling required very small silicate grains,
0.01\um\ rather than a more typical 0.1\um, assuming that the dust is
heated by the Arches stellar cluster.

Other than a location near the Quintuplet cluster, the source of the
supernovae that could excite the concentric arcs is unclear. The
Quintuplet cluster itself is an obvious candidate for the source of
supernovae producing blast waves because of its location at or near
the Arched Filaments' center of curvature and its population of
high-mass stars.  The cluster contains 9 to 13 massive WC-type
Wolf-Rayet stars \citep{liermann12}. With the age of the Quintuplet
cluster estimated at $4 \pm 1 \times 10^6$ years \citep{figer99b,
  liermann12}, the typical $10^5$ year duration of the Wolf-Rayet
phase before the star becomes a 
supernova indicates that the stars formed in a burst of star
formation.  A number of nearly coeval stars are close to becoming
supernovae, so it is plausible that three or more stars have done so
over $10^5$ years.

An argument against the Quintuplet cluster as the source of supernova
blast waves is a constraint from the cross-sky velocity difference
between the cluster and the Arches cloud. If the cluster and the
Arches cloud do not have similar proper motions the cluster will move
noticeably across the sky in the time between the supernova detonation
and the time its shell takes to expand to the radius of the
corresponding filament.  This effect is largest for filaments and arcs
with the largest radii.  For a propagation time of $1.9\times 10^5$
years to the W Arched Filament at $r = 24$\,pc, keeping the Quintuplet
centered within a tenth of the filament's radius of curvature requires
relative cross-sky velocities between the Arches cloud and the
Quintuplet cluster $\lesssim 12$\kms.  The Quintuplet cluster has a
proper motion velocity of $132 \pm 15$\kms\ along the Galactic plane
\citep{stolte14}, and its radial velocity is about +130\kms\
\citep{figer95, geballe00}.  The Arches cloud's cross-sky velocity is
unknown, and its radial velocity is $-30$\kms.  Radial velocity
differences do not affect concentricity (and are small compared with
the blast wave speeds), but such a large difference in radial
velocities suggests that the cross-sky velocities may not match
closely.

Other massive stars on the verge of becoming supernova are present
in the region near the Pistol, Sickle, and Quintuplet cluster. Some
examples include multiple LBV stars \citep{lau14}, including the
Pistol star; predictions of stars across the region in tidal streamers
torn from the the Quintuplet cluster \citep{habibi14}; and X-ray
evidence for supernovae associated with, but outside, the Quintuplet
cluster \citep{ponti15}.

The Arches cluster is likely an important source of background UV
radiation for the Arched Filaments (e.g., \citealt{erickson91,
colgan96, lang01, lang02, cotera05, hankins17}), but this younger
cluster has relatively few stars in the Wolf-Rayet phase, and is
unlikely to have produced several recent supernovae. It, along with
the Nuclear Star Cluster, is also far from the center of the bulls-eye
pattern.
  
Despite the uncertainties, the suggestion that the Arched Filaments
and other arcs in the region are byproducts of central activity merits
further exploration.  That \sgra's striking bulls-eye pattern
stretches across tens of parsecs most likely has physical meaning.
Massive stars near supernova are in the region near the Quintuplet,
and distributed supernova activity in the Galactic center is thought
to drive the fast winds from the center.  Shocks from transient winds
other than Sedov-Taylor phase blast waves are also worth considering.

\section{Summary \label{summary}} 

\cii's ability to pick out moderately excited neutral and ionized
material provides an excellent view of the Galactic center region. The
line's 158\um\ wavelength penetrates dust in the Galactic plane well,
there is little contamination from emission from clouds in the plane,
and absorption is limited to well-defined spectral bands at lower
velocities than most of the center's clouds. As one of the dominant
coolants of the ISM, the line is bright, tracing radiative and
mechanical energy deposition on the surfaces of molecular clouds.
Velocity resolution further separates physical components along the
line of sight.

Bright \cii\ highlights molecular cloud surfaces throughout the
Central Molecular Zone.  The largest molecular cloud traced by \cii\
moves in the sense of Galactic rotation, with velocity smoothly but
not linearly changing from 39\kms\ to 27\kms\ from the positive to
negative longitude edges of the field imaged here
($0.4\degr \geq \ell \geq -0.1\degr$), with a typical radial velocity
of about +28\kms.  This extended cloud forms a background for many of
the notable individual sources in the center. The cloud is prominent
in \cii, with a large, bright \cii\ region to $+\ell$ that has
received little attention in other tracers.  The surface of the next
most prominent cloud complex hosts the Arched Filaments as excitation
enhancements on a broader background, and lies at a forbidden velocity
in the sense of Galactic rotation.

Young luminous stellar clusters, winds, and individual stars ionize
the near surfaces of the extended background and Arches cloud, as well
as the Sickle region.  \cii\ intensities and lineshapes change little
across the regions including the Sickle and the top of the Arched
Filaments.  This indicates little if any interaction between the
emitting material and the magnetic fields associated with the
nonthermal filaments of the Radio Arc.

Only the core of the Galactic center's +50\kms\ molecular cloud
appears in \cii.  An upper limit on the intensity of the
[$^{13}$C\,{\sc ii}] line toward the cloud's \cii\ peak shows that the
optical depth in \cii\ is $\tau \leqslant 2.8$, and is likely below
unity.  With a velocity falling in the range where Galactic plane
absorption is strong, the +20\kms\ cloud is invisible in \cii.  The
Brick cloud obscures emission from the +28\kms\ cloud, but without
even a hint of emission in \cii\ or radio continuum tracing
ionization, the Brick cloud must be some distance in front of the
luminous clusters.

We examined the \sgra\ region and the Arched Filaments in detail.
Toward the CND, \cii\ extends from the bright edges marking the
limb-brightened southwestern and northeastern lobes.  \cii\ velocities
at peak brightness match those of the molecular emission toward the
inner the CND but remain constant with radial distance rather than
falling on the molecular rotation curves.  The extreme \cii\
velocities of $\pm 120$\kms\ correspond to circular orbits near 2\,pc,
the distance to the inner edge of the CND given the mass distribution
surrounding \sgras.  Linear structures extend from the inner edges of
the CND, possibly the result of winds from the core of the Central
Nuclear Cluster stripping and accelerating gas from the CND.  \cii\
velocity structure does not show any sign of matter flowing from
larger radii into the few parsecs around \sgras, nor any interaction
between outflows from the \sgrae\ supernova remnant and the CND.

We explored the possibility that \sgra\ region's bulls-eye pattern
traces transient excitation that adds to steadier stellar UV radiation
on the surfaces of molecular clouds.  We used a simple model to
investigate whether the Arched filaments and other arcs could be
produced at the intersection of spherical blast waves from supernovae
and cloud surfaces.  Such a model is broadly consistent with the
large-scale geometry of the region's radio continuum and X-ray
distributions.  It explains the curvature of the Arched Filaments and
their far-infrared polarization vector directions with little tuning
for particular cloud structures.  As plausibility checks, we found
that the structure and locations of radio continuum, far-IR fine
structure line emission, and dust luminosity are similar to the
corresponding tracers in the IC\,443 interaction zone of a supernova
remnant and a molecular cloud.  Massive stars near the Quintuplet
cluster are candidates for the sources of supernovae, as the region
lies near the centers of curvature of the Arched Filaments, is at the
center of the \halo, and contains a considerable population of evolved
high-mass stars. We suggest that a model involving transient winds
from central sources is worthy of further consideration and
investigation.

In addition to providing clues to the origin of the Arched Filaments,
the detailed imaging possible in far-infrared lines from the Galactic
center shows that the most intense emission in different far-IR fine
structure lines comes from different physical locations, quite
possibly tracing different excitation mechanisms.  This is a
cautionary note for using spatially averaged intensity ratios to
derive precise physical conditions within the complex regions of other
nuclei.

\begin{acknowledgments} 
  We thank the many people who have made this joint U.S.-German
  project possible, including the \grt\ and FIFI-LS teams, the SOFIA
  observatory staff, and former Science Mission Operations Directors
  E.\ Young and H.\ Yorke.  We also thank C.\ Lang for access to her
  20\,cm data; M.\ Hankins and S.\ Veilleux for stimulating
  discussions; and the anonymous referee, whose comments and
  suggestions substantially improved this paper.

  Many of our conclusions are based of published and unpublished work
  from the last 60+ years of intense interest in the Galactic center
  and its structure, and we acknowledge the progress made by the
  entire Galactic center community in understanding this complex and
  fascinating region.

  This work is based on observations made with the NASA/DLR
  Stratospheric Observatory for Infrared Astronomy (SOFIA).  SOFIA was
  jointly operated by the Universities Space Research Association,
  Inc. (USRA), under NASA contract NNA17BF53C, and the Deutsches SOFIA
  Institut (DSI) under DLR contract 50 OK 0901 to the University of
  Stuttgart. Financial support for this work was provided by NASA
  through awards 05-0022 and 06-0173 issued by USRA, by the
  Max-Planck-Institut f\"ur Radioastronomie, by the Deutsche
  Forschungsgemeinschaft (DFG) through the SFB~956 program award to
  MPIfR and the Universit\"at zu K\"oln, and by the University of
  Maryland.  DR acknowledges the financial support of DIDULS/ULS,
  through the project PAAI 2023.  This work used data from the MeerKAT
  telescope, which is operated by the South African Radio Astronomy
  Observatory, which is a facility of the National Research
  Foundation, an agency of the Department of Science and Innovation.
  AH thanks the Infrared Group of the Max-Planck-Institut f\"ur
  extraterrestrische Physik for their hospitality while some of this
  work was prepared.
\end{acknowledgments} 

\facilities{SOFIA \citealt{young12}, SOFIA/(\grt)
\citealt{risacher18}, SOFIA/(FIFI-LS) \citealt{klein14}, {\it
Herschel}/(PACS) \citealt{pilbratt10, poglitsch10}}

\software{GILDAS (\citealt{pety05, 2013ascl.soft05010G}),
  HIPE (\citealt{hipe}),
  SAOImageDS9 (\citealt{joye03, 2000ascl.soft03002S}), 
  R (\citealt{r}), Python (\citealt{python}),
  astropy (\citealt{Astropy1, astropy2, astropy3}),   
  PhotoDissociation Region Toolbox (\citealt{pound08, kaufman06}), 
  Baseline correction using splines
  (\url{https://github.com/KOSMAsubmm/kosma\_gildas\_dlc}; 
  \citealt{higgins11, kester14, higgins21})
}

\end{document}